\documentclass[10pt]{iopart}
\pdfoutput=1

\usepackage{graphicx,sidecap}
\usepackage{bm}
\usepackage{braket}
\usepackage{amssymb}
\usepackage{amsfonts}
\usepackage{mathrsfs}
\usepackage{xcolor}
\usepackage{hyperref}
\usepackage{enumerate}
\usepackage[toc,title]{appendix}
\usepackage{graphicx,epsfig,psfrag,color}
\usepackage{iopams}  
\hypersetup{colorlinks,bookmarksopen,bookmarksnumbered,
citecolor=red,
linkcolor=blue,
urlcolor=purple}
\begin{document}

\voffset=-.7cm
\textwidth=16cm

\catcode`@=11 
\renewcommand\tableofcontents{%
  \section*{\contentsname}%
  \@starttoc{toc}%
}
\catcode`@=12

\title[Disordered Contact Networks in Jammed Packings of Frictionless Disks]{Disordered Contact Networks in Jammed Packings of Frictionless Disks}
\author{Kabir Ramola} 
\address{Martin Fisher School of Physics, Brandeis University, Waltham, MA 02454, USA}
\ead{kramola@brandeis.edu}
\author{Bulbul Chakraborty} 
\address{Martin Fisher School of Physics, Brandeis University, Waltham, MA 02454, USA}
\ead{bulbul@brandeis.edu}
\date{\today}

\vspace{10pt}

\begin{abstract}
We analyse properties of contact networks formed in packings of soft frictionless disks near the unjamming transition. We construct polygonal tilings and triangulations of the contact network that partitions space into convex regions which are either covered or uncovered. This allows us to characterize the {\it local} spatial structure of the packing near the transition using well-defined geometric objects. We construct bounds on the number of polygons and triangulation vectors that appear in such packings.
We study these networks using simulations of bidispersed disks interacting via a one-sided linear spring potential. We find that several underlying geometric distributions are reproducible and display self averaging properties.
We find that the total covered area is a reliable real space parameter that can serve as a substitute for the packing fraction.
We find that the unjamming transition occurs at a fraction of covered area $A_G^{*} = 0.446(1)$. We determine scaling exponents of the excess covered area as the energy of the system approaches zero $E_G \to 0^+$, and the coordination number $\langle z_g \rangle$ approaches its isostatic value $\Delta Z = \langle z_g \rangle - \langle z_g \rangle_{\rm iso} \to 0^{+}$. We find $\Delta A_G \sim \Delta {E_G}^{0.28(1)}$ and $\Delta A_G \sim \Delta Z^{1.00(1)}$, representing new structural critical exponents.
We use the distribution functions of local areas to study
the underlying geometric disorder in the packings.
We find that a finite fraction of order $\Psi_O^* = 0.369(1)$ persists as the transition is approached.
\end{abstract}

%
%
%
%

\maketitle

\tableofcontents

\section{Introduction}
\label{Introduction}

As finite sized rigid particles are brought together by increasing their density or by compression, they undergo sharp transitions into globally rigid structures, a phenomenon known as jamming \cite{degennes_review_rmp_1999,kadanoff_review_rmp_1999, cates_bouchaud_prl_1998, bouchaud_review_2002,liu_nagel_nature_1998}. Such systems do not in general form periodic spatial patterns associated with crystalline solids, instead they display rigidity in an amorphous fashion. 
The physical properties of such amorphous solids are intimately linked with the underlying disorder in their packing.
The packing of particles into mechanically rigid structures has been of interest in diverse fields including physics, biology and geology and has wide-ranging industrial applications. 
Several characteristics of rigid particle packings have been studied experimentally and also via numerical simulations \cite{scott_kilgour_bjap_1969,jaeger_nagel_science_1992,
radjai_prl_1996,jaeger_behringer_rmp_1996,moukarzel_prl_1998,nowak_jaeger_pre_1998,
howell_behringer_prl_1999,tkachenko_witten_pre_1999,makse_prl_1999,corwin_nagel_nature_2005,
majmudar_behringer_prl_2007,olson_teitel_prl_2007,bi_nature_2011}.
Many theoretical studies have focussed on frictionless particles with hard core interactions, and much of our rigorous understanding stems from the study of such systems \cite{hansen_mcdonald_book, mccoy_book,torquato_prl_2000, torquato_stillinger_pre_2003,torquato_stillinger_rmp_2010,parisi_zamponi_rmp_2010}. 
Particles with isotropic shapes such as spheres and disks are particularly appealing in terms of their mathematical tractability, and have a long history of study in their own right \cite{parisi_zamponi_rmp_2010,aste_book_2008, bernal_nature_1959, bideau_oger_jphys_1986,aste_pre_1996,
donev_stillinger_prl_2005,donev_stillinger_pre_2005,
torquato_nature_2015}.
They have frequently been used as idealized models to study more general  problems of granular materials, glassy systems and even information theory \cite{parisi_zamponi_rmp_2010}. Although hard cores serve as a good first approximation to real particles, the incorporation of distance dependent interactions represents a non-trivial generalization which continues to be a subject of active research. A natural way to introduce such interactions is by considering particles with soft cores, with an energy dependent on the amount of overlap between particles.
Soft particle packings, which yield hard particle systems in the limit of zero overlap, have also been the subject of considerable interest \cite{durian_prl_1997,durian_pre_1997,van_hecke_review_2010,tighe_prl_2012}.
Recently such packings of soft
frictionless particles have also been realized experimentally \cite{bolton_weaire_prl_1990,brujic_makse_physicaa_2003,
dinsmore_science_2006,weeks_arxiv_2015}.  

The jamming of soft frictionless disks in two dimensions has been investigated in great detail over the last decade \cite{ohern_prl_2002,silbert_grest_landry_pre_2002,
ohern_pre_2003,wyart_nagel_epl_2005,silbert_liu_nagel_pre_2006,
henkes_chakraborty_ohern_prl_2007,henkes_chakraborty_pre_2009,
ellenbroek_pre_2009,wyart_prl_2012}. 
As soft disks are gradually compressed, they undergo a transition to a 
{\it marginally jammed state} at a well-defined protocol-dependent density. This marginal state, although at zero pressure and zero energy,
nevertheless displays peculiar spatial characteristics, such as local randomness and long range hyperuniformity \cite{torquato_stillinger_pre_2003,berthier_sollich_prl_2011,
torquato_pnas_2014,yodh_torquato_pre_2015}.
The reproducibility of several properties of marginal states, such as zero energy packing fractions and contact numbers, despite their seemingly random internal structures have led to a renewed interest in their structural properties. Several entropic arguments have been proposed to explain these characteristics of soft disks close to the jamming transition including that the marginal state is
{\it maximally random} \cite{torquato_prl_2000,torquato_pnas_2014}, or that the majority of states belong to basins that jam at the same density \cite{ohern_prl_2002,ohern_pre_2003}. 
Bulk properties of jammed soft disks have been extensively investigated and several {\it average} properties of marginally jammed disks have been well corroborated.
The average number of contacts per particle in the marginally jammed state of sphere packings is well known: $\langle z_g \rangle_{\rm iso} = 2d$ where $d$ is the dimensionality of the embedding space. 
This condition is also referred to as {\it isostaticity}.
For disks that we consider in this paper $\langle z_g \rangle_{\rm iso} = 4$. The distributions of the contact lengths, contact angles and two point correlations have also been studied in detail \cite{ellenbroek_pre_2009}. 
However despite considerable effort, several questions about jamming transitions of even simple frictionless soft disks remain unanswered \cite{van_hecke_review_2010}.
In particular, the local microstructures in packings close to the transition remain relatively less understood. Similarly, the spatial randomness of the network and the nature of the underlying disorder is an aspect that has not received much attention.
Although there is a well defined transition in the bulk properties of the system with continuously measurable quantities such as packing fraction, there are discontinuous jumps in several structural variables that are ill-defined on one side of the transition. These include network properties such as connectivities of particles, shear and pressure. It is not a-priori obvious that the two sides of the transition can be described with a single theory, and indeed by the same microscopic variables. In this context it becomes important to identify the relevant microscopic variables with which to describe the behaviour of the system near the transition.
 
\section{Summary and Overview of Results}
\label{summary_section}

\begin{figure}
\hspace{1.5cm}
\includegraphics[height=.35\columnwidth,angle=0]{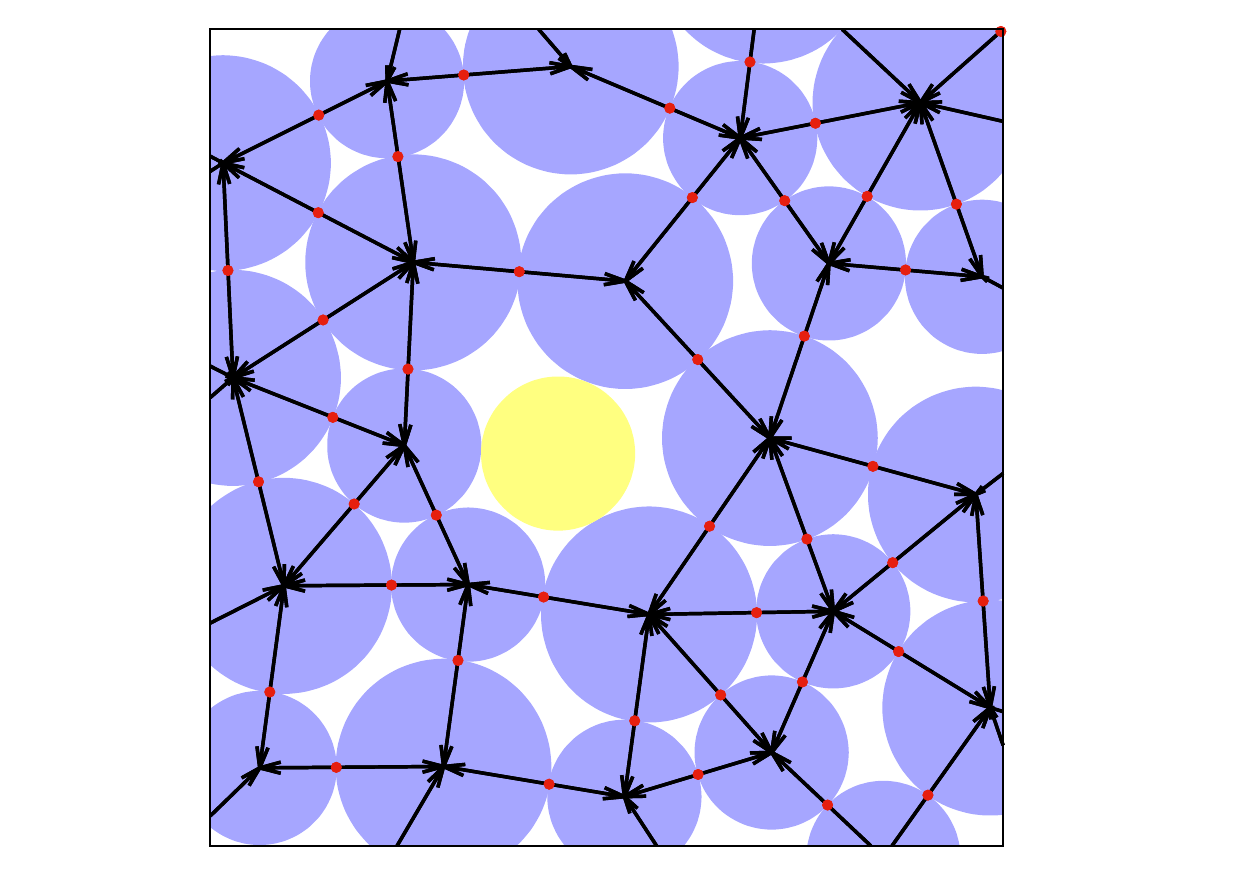}
\hspace{-1cm}
\includegraphics[height=.35\columnwidth,angle=0]{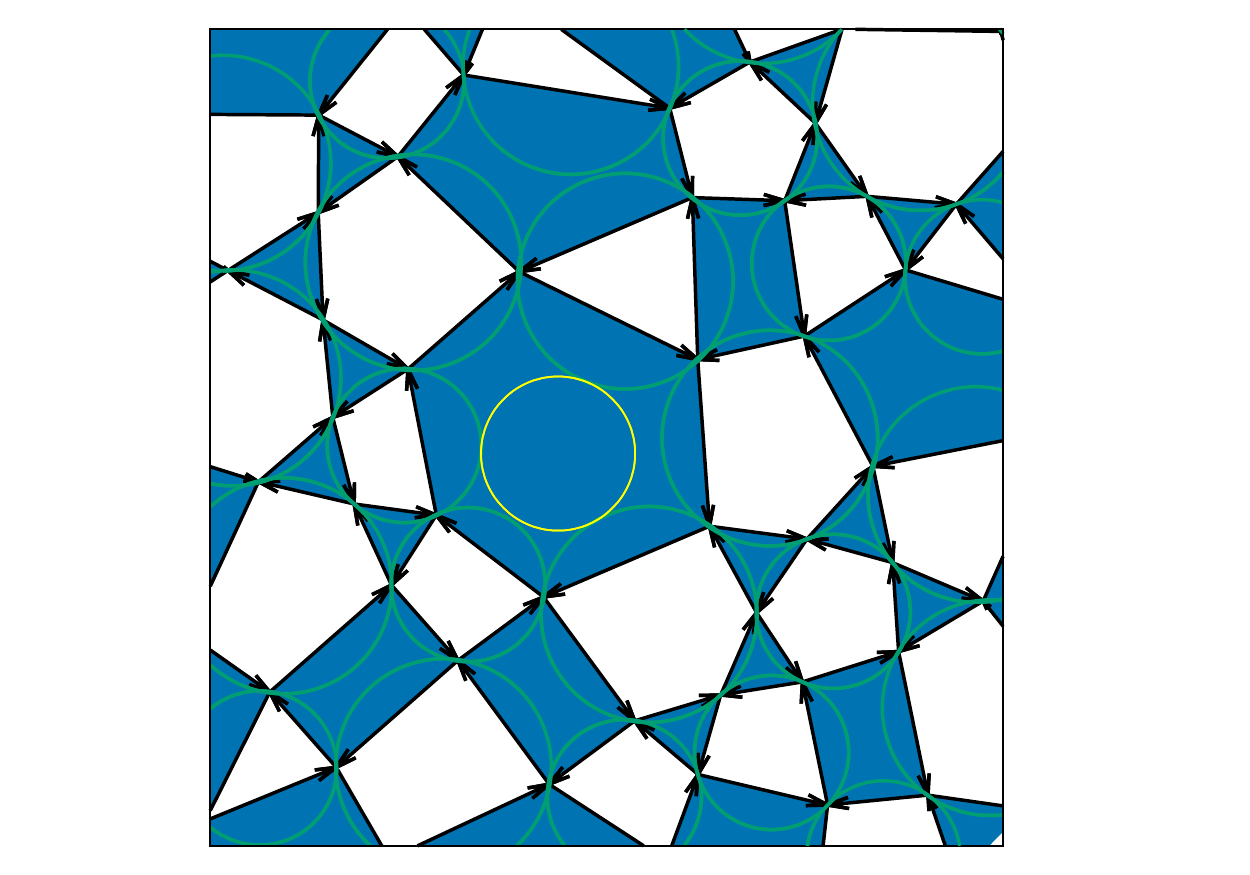}
\caption{{\bf (Left)} A section of a jammed packing of bidispersed frictionless disks. The contact points are highlighted in red. The distance vectors between the centres of grains are depicted by black (bidirectional) vectors. Particles that do not overlap with any others (rattlers) are shown in yellow. The graph  formed by the distance vectors is referred to as the ``contact network". The faces of the contact network are referred to as ``minimum cycles". The areas lying outside the grains (white) are referred to as ``voids". {\bf (Right)} The same section with ``edge vectors" cyclically connecting the contacts within each grain. The grains that are part of the contact network are depicted in green and the rattlers are shown in yellow.
This construction partitions the space into convex polygons defined as ``grain polygons"(blue) and ``void polygons" (white). Together they tesselate the entire space. The rattlers in this representation lie within the void polygons.}
\label{explanatory_figures}
\end{figure}

In this paper we study the spatial structures that arise near the well-studied jamming transition of frictionless soft disks.
We focus specifically on jammed configurations near the unjamming transition point, i.e. we approach the transition from mechanically stable packings with decreasing energies. 
The main results of this paper can be summarized as follows. We introduce a representation that assigns convex polygonal areas to the disks and {\it also} to the voids (see Fig. \ref{explanatory_figures}). We also assign a measure to this polygonal construction using a Delaunay triangulation of the underlying network. This allows us to study the structural properties of both voids and disks and accurately measure their statistics. Although the underlying degrees of freedom have complicated joint probability distributions arising from constraint satisfaction, we find that they display self averaging behaviour, such that system averages are equivalent to ensemble averages. This allows us to measure reproducible distribution functions of these individual quantities.
We find that the total area occupied by the grain polygons serves as a reliable real space parameter that displays non-trivial scaling as the transition is approached. We perform large scale simulations to measure scaling exponents associated with this area as the transition is approached.
We then distribute this total area into microscopic variables by assigning well-defined areas to each contact and find this works well as a local field. Finally we use this construction to study the nature of the disorder that arises in such packings. Our study relies on the special convexity properties that appear in frictionless force balanced systems (see Section \ref{convexity_section}). This
highlights the difference between the contact networks that emerge in frictional and frictionless systems. We are not aware of any other real space representation where such a difference is manifest. 

The paper is organised as follows. 
In Section \ref{Generating_jammed_section} we describe the potential energy function and define the jammed configurations we focus on in our study.
In Section \ref{Polygonal_representation_section} we introduce two representations of the disk packing referred to as the ``polygonal" and ``triangulation" graphs. This allow us to accurately characterise the spatial properties of jammed disk packings.
In Section \ref{network_properties_section} we characterise the total number of voids, contacts and triangulation vectors that arise in such packings.
We derive conservation laws for the triangulation graph that allows us to assign the triangulation vectors uniquely to the local polygonal objects.
We use these network properties to characterise the spatial degrees of freedom of the system and use these geometrical degrees of freedom to construct an isostaticity condition.
In Section \ref{distribution_functions_section} 
we perform large scale simulations to measure statistics of these contact networks, focussing on the lengths of contact vectors, their relative angles and the local areas of the polygons.
In Section \ref{grain_area_section} we provide evidence that the total grain area can be used as a reliable parameter with which to describe the unjamming transition. 
In Section \ref{measures_of_order_section} 
we use the distribution functions of the local areas to study
the underlying geometrical randomness in the packings.
We find that a finite fraction of order persists as the transition is approached. We also find that the excess order in the system displays non-trivial scaling as the transition is approached.
Finally, in Section \ref{correlation_section} we measure particle and contact correlations that we use to estimate the length scales over which the jammed packings display disorder.

Although we focus specifically on frictionless disks, our methods can be readily applied to frictional systems and other convex particles as well.
In this paper, we refer to the particles interchangeably as disks or ``grains". We use capitalized letters to represent global quantities and lower case letters to represent local properties.

\section{Jammed Configurations}
\label{Generating_jammed_section}
\begin{figure}
\hspace{1.5cm}
\includegraphics[height=.35\columnwidth,angle=0]{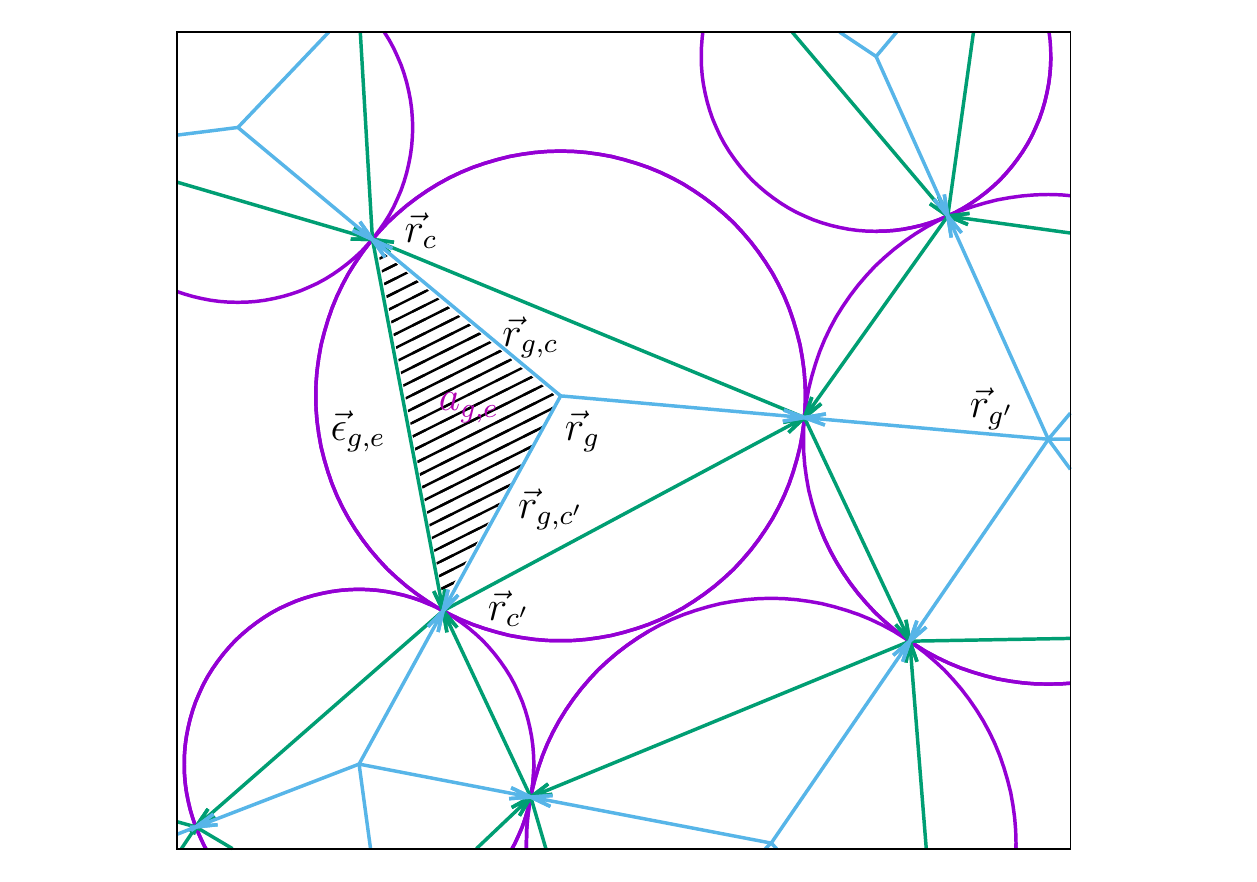}
\hspace{-1cm}
\includegraphics[height=.35\columnwidth,angle=0]{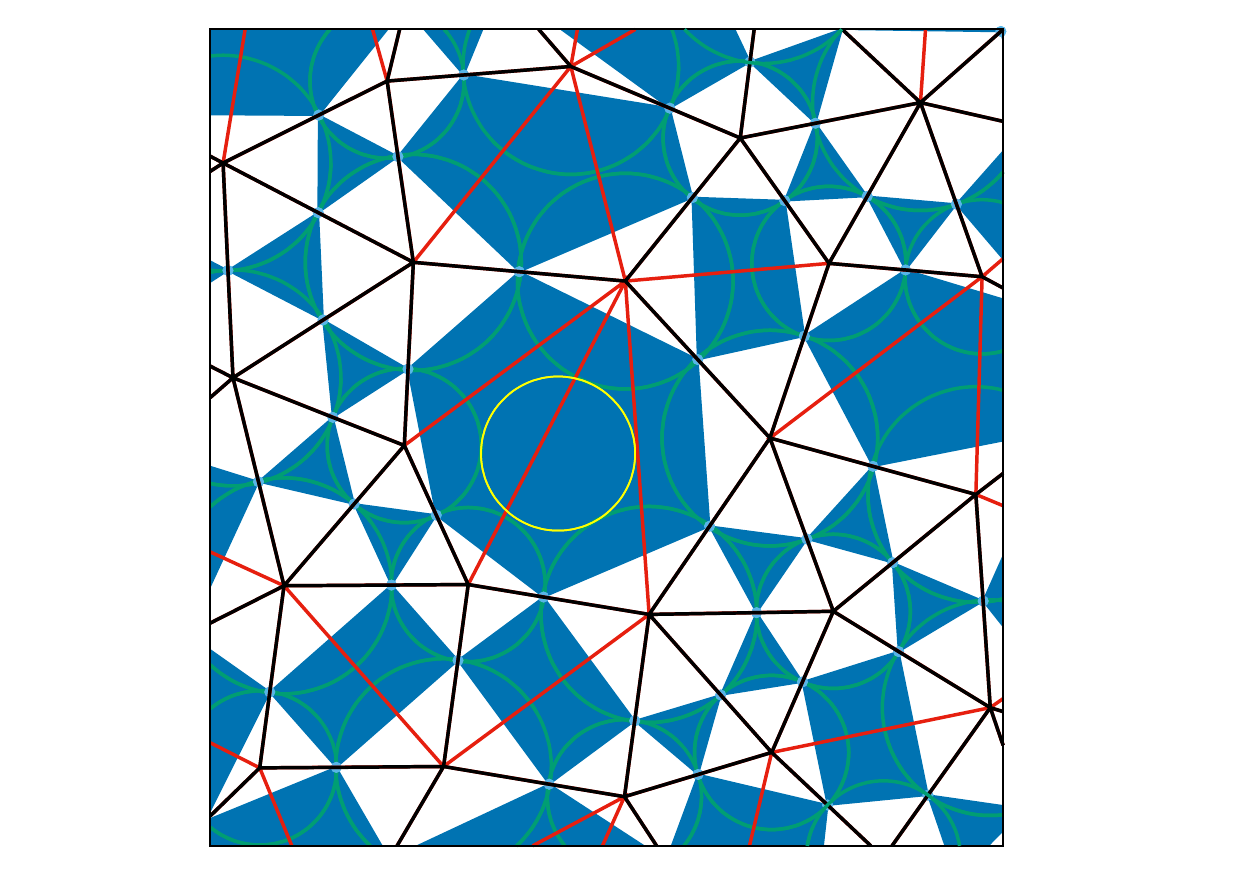}
\caption{{\bf (Left)} The labeling convention. The positions of the centers of the grains are represented by the set of vectors $\{ \vec{r}_{g} \}$ with associated radii $\{\sigma_g\}$. 
The positions of the contact points are represented by the set of vectors $\{ \vec{r}_{c} \}$ where $c = 1,2,3 ... N_C$, the total number of contacts. The contact vectors belonging to every grain $g$ are represented by $\vec{r}_{g,c}$ where $c = 1,2,3...z_g$, the connectivity of the grain.
The edges associated with every grain $g$ are represented by $\{ \vec{\epsilon}_{g,e}\}$ with $e = 1,2,3 ... z_g$. The edges within each grain circulate in an anti-clockwise direction. The triangle formed by the points $(\vec{r}_{g}, \vec{r}_{c}, \vec{r}_{c'})$ (shaded area) is labelled by the unique edge $\vec{\epsilon}_{g,e}$ and is referred to as an ``edge triangle" with an associated area $a_{g,e}$.
{\bf (Right)} The jammed configuration of Fig. \ref{explanatory_figures} along with the associated polygonal construction and Delaunay triangulation. The vectors coinciding with real contacts are depicted in black and the ``fictitious contacts" are shown in red.}
\label{system_figure}
\end{figure}

We begin by defining an energy function for a given configuration of disks.
A configuration is fully specified by the set of positions of the centres of grains $\{ \vec{r}_{g} \}$ and their associated radii $\{\sigma_g\}$. 
The total energy of the system is given by a sum of two body interactions $V(\vec{r}_{g, g'})$, where $\vec{r}_{g,g'} = \vec{r}_{g'} - \vec{r}_{g}$ is the distance between the disks $g$ and $g'$. This interaction potential is modelled as a one-sided linear spring repulsion potential of the form
\begin{equation}
V(\vec{r}_{g, g'}) = \frac{1}{2} \left(1-\frac{|\vec{r}_{g,g'}|}{\sigma_{g,g'}}\right)^2 \Theta\left(1-\frac{|\vec{r}_{g,g'}|}{\sigma_{g,g'}}\right),
\label{potential_energy_functional}
\end{equation}
where $\Theta$ is the Heaviside function and $\sigma_{g, g'} = \sigma_{g} + \sigma_{g'}$ is the sum of the undistorted radii of disks $g$ and $g'$. The total number of grains is referred to by $N_G$. When quoting $N_G$ from simulation data, we follow the convention that all grains used in the simulation are counted, regardless of whether they are a part of the contact network in the final jammed state. In computing all other system quantities, $N_G$ refers only to the grains that are part of the rigid structure, i.e. $N_G \equiv N_G - N_R$, where $N_R$ is the number of ``rattlers", particles that are not in contact with any of the others (see Fig. \ref{system_figure}).
The total energy per grain of the system is then given by
\begin{equation}
E_G = \frac{1}{N_G} \sum_{(g,g')} V(\vec{r}_{g, g'}),
\label{full_system_energy}
\end{equation}
where the sum is taken over all pairs $(g,g')$, with $g \ne g'$.
The potential energy function therefore acts as a tuning parameter that defines a distance to the unjamming transition located at precisely $E_G = 0$. In this paper we approach the transition from finite positive energies $E_G \to 0^{+}$.
A configuration is defined as jammed if and only if $E_G > 0$ {\it and} 
\begin{equation}
\sum_{g' \ne g} \frac{\partial V(\vec{r}_{g,g'})}{\partial \vec{r}_g} = 0 ~~\forall~~ g.
\label{force_balance_equation}
\end{equation}
Eq. (\ref{force_balance_equation}) is simply a statement of {\it force balance} for every grain $g$. Two disks are in contact if $|\vec{r}_{g,g'}| < \sigma_{g,g'}$.
We refer to the collection of distance vectors $\{ \vec{r}_{g,g'} \}$ between grains in contact as the ``contact network". The force balance condition ensures that configurations with finite energies possess a system spanning contact network.
In this paper we only consider configurations that are jammed.
We consider systems with periodic boundary conditions in both directions. This boundary condition makes the embedding plane a genus $1$ torus with an Euler characteristic $\chi = 0$. We will use this property in Section \ref{network_properties_section}.

\section{Contact Network}
\label{Polygonal_representation_section}
We next introduce two representations of the disk packing generated using the network of contacts that form between the disks. We generate a ``polygonal graph" by partitioning the two dimensional space into polygonal tilings of grains and voids, and a ``triangulation graph" by a Delaunay triangulation of locally convex sections of the contact network. Recently
Voronoi tesselations have proved fruitful in understanding the local spatial structures of packings \cite{morse_corwin_prl_2014}. Our procedure differs from the Voronoi tesselation in that it is able to assign well defined sections of space to grains as well as to voids. 

The packing of disks in the two dimensional plane can naturally be considered as a decomposition of space into regions that lie within the circular areas associated with the disks and regions that lie outside. We refer to these exterior regions as ``voids".
In jammed packings, these voids are completely enclosed by the areas of grains surrounding them, and in general have a complicated, non-convex shape (see Fig. \ref{explanatory_figures}). 
The total number of such voids is referred to by $N_V$. 
In order to characterize these voids we can use the positions of the surrounding grains to build  circuits around them. This is done by constructing a circuit of distance vectors with vertices at the centres of the grains surrounding each void $v$ as $\mathcal{M}_v = \{\vec{r}_{g,g'}, \vec{r}_{g',g''}, ...  \vec{r}_{g''',g} \}$ (see Fig. \ref{explanatory_figures}). These circuits are also referred to as {\it minimum cycles} of the contact network since they represent the shortest possible loops in this network. Each minimum cycle is uniquely associated with a void, and completely encloses the non-circular shaped space formed between the disks.

\subsection{Contact Vectors and Edges}
Using the positions and radii of the grains in a given configuration we can next construct the contacts between the grains. Although the contacts occur as finite regions of overlap between grains, we assign them to unique points in space.
The positions of the contact points are represented by the set of contact positions $\{ \vec{r}_{c} \}$. We use the convention 
\begin{equation}
\vec{r}_{c} = \vec{r}_g + \frac{\sigma_g}{\sigma_g + \sigma_g'}\left( \vec{r}_{g'} - \vec{r}_{g} \right),
\label{contact_positions_equation}
\end{equation}
where $c$ is the contact between grains $g$ and $g'$. Here $c = 1,2,3...N_C$ where $N_C$ is the total number of contacts in the system.
In this convention every contact is counted twice, once for each grain.
The convention  in Eq. (\ref{contact_positions_equation}) also ensures that configurations that are related by a simple expansion or contraction of all disks have the same contact network. Using the positions of contacts belonging to every grain, we can assign ``contact vectors" to each grain as
\begin{equation}
\vec{r}_{g,c} = \vec{r}_c - \vec{r}_g,
\end{equation}
where $g$ refers to the grain index and $c$ refers to the contact number associated with the grain, $c = 1,2,3...z_g$. The number of contacts $z_g$ is also referred to as the ``connectivity" of the grain. 
Every contact vector has two degrees of freedom which can be parametrized by its length $|r_{g,c}|$ and its angle in relation to the $x$-axis $\theta_{g,c}$.
The system can then be parametrized in terms of the contact vectors, although this is an overcounting of the degrees of freedom.  The contact vectors must satisfy constraints in order for them to be a valid packing. We discuss these constraints in detail in Section \ref{coordination_section}.

\subsection{Semi-Circle Condition and Convexity}
\label{convexity_section}

The normality of forces in the system of frictionless disks leads to special convexity properties of the contact network. The force balance condition on each grain disallows any configuration in which contacts lie on {\it only} one semi-circle of a disk, since this would generate a net force along the direction normal to this semi-circle. We refer to this statement as the 
``semi-circle condition". This then ensures that the relative angles between the contact vectors obeys
\begin{equation}
\theta_{g,c'} - \theta_{g,c} < \pi,
\end{equation}
where $(c,c')$ is taken cyclically within each grain.
Using this property, we can deduce that each minimum cycle surrounding a void forms a convex polygon. Therefore the contact network partitions space into {\it locally convex} regions, with each of these regions being uniquely associated to a void. The partitioning of space into locally convex regions greatly simplifies the characterization of the underlying structure of the packing and allows us to construct convex tilings associated uniquely with grains and also with the voids.
It is important to note that there is no such semi-circle condition for frictional systems. The network construction therefore highlights a crucial difference between frictionless and frictional systems.

\subsection{Grain Polygons and Void Polygons}
\label{grain_void_section}

Using the contact vectors, we can next build a collection of edges $\{ \vec{\epsilon}_{g,e}\}$ defined as
\begin{equation}
\vec{\epsilon}_{g,e} = \vec{r}_{g,c'} - \vec{r}_{g,c}. 
\end{equation}
where the edges connect pairs of contacts $(c,c')$ belonging to the grain $g$ in a cyclic manner based on their angles (see Fig. \ref{system_figure}).
Here $e$ is the edge number associated with the grain, $e = 1,2,3...z_g$. We use the convention that the edges circulate in an anti-clockwise manner within each grain. Every contact belonging to a grain therefore has a unique edge associated with it, namely the outgoing edge vector at each contact. The edges associated with the grains naturally form a $z_g$-sided polygon. We refer to these embedded polygons as ``grain polygons". Since we are dealing with convex disks, the associated polygons are also convex and hence have an associated convex area $a_g$. The total area covered by the grain polygons is denoted by $A_G$. We then have
\begin{eqnarray}
\nonumber
a_{g,e} = \frac{1}{2} \left( \vec{r}_{g,c} \times \vec{r}_{g,c'} \right),\\
{\rm with}~~a_g = \sum_{e} a_{g,e} ~~{\rm and}~~ A_G = \sum_{g} a_{g}.
\label{grain_area_eq}
\end{eqnarray}
where the edge $e$ connects the contacts $(c,c')$. We refer to the individual areas $a_{g,e}$ as ``edge triangles", since they are triangular in shape and are uniquely associated with an edge $e$ (see Fig. \ref{system_figure}). We measure the statistical properties of these edge triangles in Section \ref{measures_of_order_section}. 

Next, we can also use the contact vectors to construct ``void polygons". These are created using the edges surrounding each void in the same manner as the grain polygons and have an associated area $a_v$. For fricitionless packings of disks, the convexity of the circumscribing minimum cycle enclosing every void also ensures the convexity of the void polygon. The edges associated with the void polygons circulate in the clockwise direction. Once again we can define a void connectivity $z_v$ as the total number of edges associated with the void. An important property of the void polygons is that they circumscribe the voids between the disks, which has a non-convex shape, with a convex polygon. 
For marginally jammed disks, the void polygon is precisely the {\it convex hull} of the non-circular area of the void.
This makes the characterization of the shape and size of individual voids easier. 
We note that our void construction is similar to the circulating currents of Ball and Blumenfeld \cite{ball_blumenfeld_prl_2002}, however their assignment of grain areas is different. Our construction is aimed at introducing a grain area parameter that serves as a substitute for the packing fraction, which can then be used to test scaling properties of the system.
We refer to the graph $\mathcal{G}_P = (\{\vec{r}_c\}, \{\vec{\epsilon}_{g,e} \})$ with vertices at the position of contacts and edges formed by the edge vectors as the ``polygonal graph".

\subsection{Delaunay Triangulation and Fictitious Contacts}
\label{delaunay_fictitious_section}

In order to assign a measure to our network, we use the positions of the grains to construct a triangulation of the two dimensional space. As we have seen, for frictionless disk packings, the contact network naturally breaks up the space into convex minimum cycles enclosing the voids. We can then individually triangulate each cycle with the distance vectors $\vec{r}_{g,g'}$ that form the convex hull of each cycle, and extra vectors $\vec{r}^{f}_{g,g'}$ traversing through the minimum cycles. We refer to these extra triangulation vectors as ``fictitious contacts" since they mostly represent disks that are almost in contact. It should be noted that the fictitious contacts are infact distance vectors in our convention. The statistics of these fictitious contacts are of interest to us since they provide significant information about how the contact network is embedded in space. 
The union of the individual triangulations then yields a global triangulation $\mathcal{G}_T = (\{\vec{r}_g\}, \{\vec{r}_{g,g'},\vec{r}^{f}_{g,g'} \})$ with vertices at the position of grains and edges formed by the grain distances and fictitious contacts.
We refer to $\mathcal{G}_T$ as the ``triangulation graph".

The natural triangulation to use within each minimum cycle is the Delaunay triangulation \cite{deberg_book_computational_geomtery}, which maximizes the minimum angle among all possible triangulations. This forms the least scalene triangulation possible and is convenient for developing discrete calculus frameworks for microscopic properties of granular systems \cite{degiuli_pre_2011,degiuli_thesis_2013,degiuli_epl_2014}.
Since the union of Delaunay triangles yields the convex hull \cite{deberg_book_computational_geomtery}, this ensures that the distance vectors are a subgraph of the triangulation of each minimum cycle. 
Therefore the contact network forms a subgraph of the triangulation graph.
This construction is unique since each minimum cycle contains points that are not collinear, ensuring the uniqueness of the individual triangulations. It is important to note that the union of the individual Delaunay triangulations does not necessarily yield the global Delaunay triangulation obtained by simply triangulating the vertices of grain positions. 
For monodisperse disks, this union does indeed yield the global Delaunay triangulation. To show this it is sufficient to prove that two disks in contact are connected by a global Delaunay edge. This can be proved by recognizing that the Delaunay triangulation is simply the adjacency graph of the Voronoi tesselation, i.e. if two points share an edge in a Voronoi tesselation, they will be connected by a Delaunay edge. The voronoi area associated with each vertex $\vec{r}_g$ is defined as all points $\vec{r}$ such that $|\vec{r} - \vec{r}_g|  < |\vec{r} - \vec{r}_{g'}| ~\forall~ g'$. 
For non-overlapping disks, the associated circular areas therefore lie within their Voronoi areas. At each overlap the Voronoi area of two disks in contact are modified by the amount of overlap, and therefore are adjacent in the Voronoi tesselation, leading to them being connected by a Delaunay edge. This property therefore breaks down if one considers highly stressed states with more than two disks in overlap.

It is easy to show that for bidispersed systems with particle radii $\sigma_A$ and $\sigma_B$ with $\sigma_A/\sigma_B < 1 + \sqrt{2}$, as $E_G \to 0^+$ the global Delaunay triangulation coincides with the union of Delaunay triangulations of minimum cycles. This is true for the bidispersed case with diameter ratio $1:1.4$ that we simulate.
In the case of highly polydisperse systems the global Delaunay triangulation is no longer the best choice, however our construction using the union of minimum cycles still holds.
To deal with such cases several generalizations of Voronoi tesselations have been proposed in the literature, the best studied of which is the radical Voronoi tesselation  \cite{gellatly_finney_jncs_1981}. 
Using this construction, the argument for the monodispersed case can easily be generalized to systems with varying sizes of disks.

\section{Network Properties}
\label{network_properties_section}
The polygonal graph and the triangulation graph are both representations of the same underlying system. However, they have different properties. We list some of the properties for each graph below.

\subsection{Polygonal Graph}

\begin{itemize}

\item
The grain polygons and void polygons form a {\it bipartite graph}, i.e. voids are connected to only grains through their edges, and vice versa. This can also be stated alternatively as the {\it adjacency graph} of grains and voids forms a bipartite network. For the hexagonal close packed structure this adjacency graph is simply the dice lattice.

\item
The network is {\it space filling}, i.e.

\begin{equation}
A_G + A_V = 1,
\end{equation}
where $A_G $ and $A_V$ are the total areas associated with the grain polygons $A_G = \sum_{g} a_g $ and the void polygons $A_V = \sum_{v} a_v$ respectively. This property differs crucially from other measures of packing fraction used in the literature, that include the excess volumes of overlaps between disks. 

\item
Every node has exactly four neighbours ($z = 4$). Every contact has two incoming edges and two outgoing edges since each contact belongs to two grains. 

\item
The network is {\it planar}, i.e. none of the edges formed by the contact vectors cross each other. This property allows us to use Euler's theorem to understand the properties of the graph.

\end{itemize}

\subsection{Triangulation Graph}

\begin{itemize}
\item
The contact network forms a subgraph of the triangulation graph. This is true by construction (see Section \ref{delaunay_fictitious_section}).

\item
The network is {\it planar}, i.e. none of the network vectors cross each other. This property follows directly from the fact that we only consider two-disk overlaps as valid and the properties of the Delaunay triangulation \cite{deberg_book_computational_geomtery}.

\item
{\it Triangulation Property:}
Every triangulation has the following general property \cite{deberg_book_computational_geomtery}:
Let $P$ be a set of $n$ points in the plane, not all collinear, and let $k$ denote the number of points in $P$  that lie on the boundary of the convex hull of $P$.  Then any triangulation of $P$ has:
$2n-2-k$ triangles and
$3n-3-k$ edges.
We will use these properties to construct conservation laws in Section \ref{conservation_section}. 


\end{itemize}

\begin{table}
\hspace{3cm}
    \begin{tabular}{| l | l | l |}
    \hline
     & Polygonal Graph & Triangulation Graph\\ \hline \hline
     $V$ & $N_C/2$ & $N_G$\\ \hline
	 $E$ & $N_C$ & $N_C/2 + N_F$\\ \hline
     $F$ & $N_G + N_V$ & $N_T = 2 N_G$\\ \hline
     Euler & $N_C/2 = N_G + N_V$ & $N_C/2 = 3 N_G - N_F$ \\ \hline
    \hline
  	\end{tabular}
  	\caption{Properties of the polygonal and triangulation graphs. $V,E$ and $F$ refer to the number of vertices, edges and faces respectively. The planarity of these graphs ensures that they obey the Euler identity: $V - E + F = \chi$, where $\chi$ refers to the Euler characteristic of the embedding plane. For the toroidal boundary conditions that we consider, $\chi = 0$. The Euler identity for the triangulation graph can be used to compute the total number of fictitious contacts in the system using the triangulation condition $N_T = 2 N_G$.}
\label{table1}
\end{table}

\subsection{Conservation Laws}
\label{conservation_section}

We next consider the intersection of the polygonal and triangulation graphs. In doing so we can assign distinct regions of the triangulation graph to regions of the polygonal graph. The grain distances can be uniquely associated with the grain polygons as each of these vectors lies completely within two grain polygons in contact. The question of assigning fictitious contact vectors uniquely to regions of the polygonal graph is more subtle. We show below that these fictitious contact vectors can be uniquely associated with the voids.

To do so we derive local conservation laws associated with every void polygon.
The semi-circle condition stated in Section \ref{convexity_section} necessitates that the centre of each disk lie {\it within} its grain polygon. We can then introduce a topological winding number for every grain polygon, defined as
\begin{equation}
w_g = \frac{1}{2 \pi} \sum_{(c,c')} \left(\theta_{g,c'} - \theta_{g,c} \right),
\end{equation}
where the sum $(c,c')$ is taken cyclically around every grain.
The semi-circle condition can then be stated alternatively as: the winding number of each grain polygon is equal to $1$.
This property breaks down for frictional packings since contacts on a single side of a disk can be stabilized by tangential forces. The winding numbers therefore add discrete structural degrees of freedom to frictional systems.

Next, since the enclosing minimum cycle for every void is convex and because the distance vectors and the fictitious contacts within each cycle together form a triangulation, the remaining triangulation vectors within this circuit must be fictitious contacts, traversing through the void polygon. We therefore associate these fictitious contacts with the void polygon. This association is unique as every fictitious contact has a single corresponding void polygon.
Then using the general property of triangulation, and since the contacts making up the void all lie on the convex hull of the void,
we arrive at the following {\it local conservation law} for every void polygon:
\begin{equation}
n_f = z_v - 3,
\label{fictitious_conservation}
\end{equation}
where $n_f$ is the number of fictitious contacts associated with every void polygon. Similarly we can associate all the triangles formed by the fictitious contacts within the convex hull uniquely to each void.
Once again following the circuit of distance vectors enclosing the void polygon and using the general triangulation condition, leads to
\begin{equation}
n_t = z_v - 2,
\label{triangle_conservation}
\end{equation}
where $n_t$ is the number of triangles associated uniquely with every void polygon. 

As a check, we can sum these local properties over the entire system to yield the Euler conditions. For the toroidal boundary conditions we consider in this paper, the Euler characteristic is $\chi = 0$.
The Euler conditions for the polygonal and triangulation graphs are summarized in Table \ref{table1}.
Summing Eq. (\ref{fictitious_conservation}) leads to
\begin{equation}
N_F = \sum_{z_v = 3}^{\infty} (z_v - 3)n_v(z_v).
\end{equation}
where $N_F$ is the total number of fictitious contacts, and  $n_v(z_v)$ is the number of void polygons with $z_v$ sides in a given configuration. In the above summation we have disregarded all voids with $z_v < 3$. This can occur if a disk has only two contacts, leading to the contacts being connected by a single void with a vanishing area, i.e. exact alignment of forces. This is atypical in a general packing and disfavoured entropically, we do not consider such packings as valid in our analysis. 
Since $\sum z_v n_v(z_v) = N_C$ (the total number of contacts), we have
\begin{equation}
N_F = N_C - 3 N_V.
\label{fictitious1}
\end{equation}
The Euler condition for the triangulation graph yields (see Table \ref{table1}) 
\begin{equation}
2 N_F + N_C = 6 N_G.
\label{Euler_triangulation}
\end{equation}
This property was also derived for the monodisperse case in \cite{degiuli_thesis_2013}. Using Eqs. (\ref{fictitious1}) and Eqs. (\ref{Euler_triangulation}) leads to
\begin{equation}
N_F = 2 N_G - N_V.
\label{fictitious2}
\end{equation}
Eq. (\ref{fictitious1}) and Eq. (\ref{fictitious2}) together yield the Euler condition for the polygonal graph $N_C = 2 N_G + 2 N_V$.
Dividing this Euler condition by $2 N_C$ gives the duality relation for the average connectivities of the grain and void polygons
\begin{equation}
\frac{1}{\langle z_g \rangle} + \frac{1}{\langle z_v \rangle} = \frac{1}{2}.
\label{duality_relation}
\end{equation}
Finally, summing the triangle conservation law in Eq. (\ref{triangle_conservation}) over the entire system yields the total number of triangles $N_T$ as
\begin{equation}
N_T = N_C - 2 N_V,
\label{triangles1}
\end{equation}
which when combined with the Euler condition for the polygonal graph, yields the Euler condition for the triangulation graph.
Conservation laws such as the one derived in this section could  prove useful in building microscopic models for frictionless networks.

\subsection{Coordination and Isostaticity}
\label{coordination_section}

Since the number of fictitious contacts is a strictly positive quantity, this can be used to construct bounds on the total number of voids and contacts in the system. Eq. (\ref{fictitious1}) leads to the trivial bound on the average void coordination of marginally overlapping disks, $\langle z_v \rangle \ge 3$. Eq. (\ref{fictitious2}) leads to the non-trivial bound 
\begin{equation}
N_V \le 2 N_G.
\end{equation}
Once again imposing strict positivity on $N_F$ from Eq. (\ref{Euler_triangulation}) we arrive at 
\begin{equation}
\langle z_g \rangle \le 6.
\end{equation}
The two conservation laws Eqs. (\ref{fictitious1}) and (\ref{fictitious2}) are saturated when $N_F = 0$, leading to $N_V = 2 N_G$ and $N_C = 3 N_V$ and consequently to $\langle z_g \rangle = 6$ for toroidal boundary conditions. 
An important consequence of the local conservation law is that since $n_f \ge 0$, this requires that {\it every} void is exactly $z_v = 3$ when this bound is saturated.
This is true for the hexagonally close packed structure.
We also note that this bound is independent of polydispersity, since the only conditions we imposed on the packing was frictionless force balance. This agrees with previous bounds on maximum coordination numbers for disk packings \cite{aste_pre_1996}.

The next question we address is the lowest coordination available to the system.
It has been noticed in several studies that marginally jammed packings ($E_G \to 0^{+}$) are exactly isostatic \cite{ohern_prl_2002, ohern_pre_2003}. 
This can be argued from the fact that there are no special conservation laws for the forces, however a sound theoretical argument for this statement is still missing.
The standard isostaticity argument for frictionless systems can be summarized as follows: the basic degrees of freedom in the system are the forces that determine whether a system is rigid, i.e. force balanced. These vector forces can be decomposed into two scalar components $f_N$ and $f_T$, the normal and tangential components at each contact respectively. For frictionless systems $f_T = 0$ identically at every contact. Since each force is uniquely associated with a contact, the total number of contact forces in the system is equal to the total number of contacts $N_C$. Next, since Newton's third law has to be satisfied at each contact, this reduces the total number of free variables by a factor of $2$, and hence the number of degrees of freedom $N_{\rm dof} = N_C/2$. The constraints of force balance are determined at every grain, these are {\it vector} constraints at each grain, yielding the total number of constraints $N_{\rm constraints} = N_G \times d$, where $d$ is the dimension of system. At isostaticity, the number of degrees of freedom exactly matches the number of constraints, and hence $\langle z_g \rangle = 2 d$. 
This argument, although correct, requires an implicit knowledge of the angles of the forces in order to move from the scalar variable $f_N$ to the vector contraints at each grain. It also invokes constraints in real space to impose constraints on the forces \cite{phillips_jncs_1979}. In addition, since this is essentially a mean field analysis, the argument ignores loop constraints \cite{thorpe_jncs_1983}. This argument also fails to predict isostatic values for systems with other types of convex particles.
An interesting aspect of isostaticity is that it is independent of the force law for frictionless disks and spheres, hinting at a more basic geometric origin of the isostatic condition. 

We can construct an isostaticity argument for frictionless disk packings using {\it only} the spatial degrees of freedom from the polygonal graph as follows. At $E_G = 0$ the distances between the contacts and the centres of the disks are completely determined. Therefore the independent degrees of freedom in the system are the angles of each contact vector. For disk packings, the contact angles for a contact between disks $g$ and $g'$ are related by
\begin{equation}
\theta_{g,c} = 2 \pi - \theta_{g',c}.
\end{equation}
Using the above condition and accounting for one global rotational degree of freedom gives us the total degrees of freedom of the system
\begin{equation}
N_{\rm dof} = \frac{N_C}{2} - 1.
\label{number_of_dof}
\end{equation}
Given the angles in the system, the edge lengths are completely determined by
\begin{eqnarray}
\nonumber
\epsilon^x_{v,e} \equiv \epsilon^x_{g,e} = \sigma_{g}(\cos \theta_{g,c'} - \cos \theta_{g,c}),\\
\epsilon^y_{v,e} \equiv \epsilon^y_{g,e} = \sigma_{g}(\sin \theta_{g,c'} - \sin \theta_{g,c}),
\label{edges_from_angles}
\end{eqnarray}
where the edges can be uniquely assigned to a grain $\vec{\epsilon}_{g,e}$ or the corresponding void $\vec{\epsilon}_{v,e}$.
This mapping from angles to edge vectors is unique and invertible, i.e. given a configuration of edge vectors $\{ \vec{\epsilon}_{g,e}\}$, the contact angles $\{\theta_{g,c}\}$ are completely determined.
However, these degrees of freedom are not all independent. In order for the set of edge vectors to be a valid packing they must satisfy loop constraints within each grain and void.
The constraints for every grain, i.e. the edge vectors within each disk form a cyclic polygon, are automatically satisfied by Eq. (\ref{edges_from_angles}). 
Next, we need to construct a set of independent constraint equations, such that a combination of any of these equations do not yield any of the others, that determine all the constraints in the system.
For frictionless systems, these equations can be easily seen to be the loop constraints for every void polygon
\begin{eqnarray}
\nonumber
\sum_{e} \epsilon^x_{v,e} = 0 ~~\textmd{for every void polygon $v$},\\
\sum_{e} \epsilon^y_{v,e} = 0 ~~\textmd{for every void polygon $v$}.
\end{eqnarray}
It should be noted that these equations represent {\it all} the spatial constraints in the system, since larger loop constraints can be built by summing the constraints of the individual void polygons enclosed by such a loop. Such sets of constraint equations have been recognized in the literature \cite{satake_mom_1993,ball_blumenfeld_prl_2002, blumenfeld_prl_2004} and were recently used to determine forces from the positions of jammed packings of soft disks \cite{procaccia_prl_2016}.
Clearly each of these loop constraints is independent of the others, since each void polygon has distinct edges. 
In this respect, this is an ``entropic" argument, i.e. we demand that the void polygons do not possess special conservation laws amongst themselves that renders some constraints unnecessary. 
We expect this to be true for states prepared by an unbiased sampling of all available solutions of Eq. (\ref{force_balance_equation}).
Hence the total number of {\it independent} constraints is 
\begin{equation}
N_{\rm constraints} = 2 N_V.
\label{number_of_constraints}
\end{equation}
For the packing to be valid, it should satisfy all possible geometric constraints. This occurs if and only if $N_{\rm dof} \ge N_{\rm constraints}$. 
Using Eqs. (\ref{number_of_dof}) and (\ref{number_of_constraints}), we arrive at the Maxwell criterion 
\begin{equation}
N_C \ge 4 N_V + 2.
\label{isostatic_inequality}
\end{equation}
Next, using the Euler condition for the polygonal graph, we arrive at the lower bound on the number of voids in the system
\begin{equation}
N_V \ge N_G -1.
\label{void_bound}
\end{equation}
which combined with Eq. (\ref{fictitious2}) leads to a bound on the total number of fictitious contacts in the system
\begin{equation}
N_G + 1 \ge N_F.
\label{fictitious_bound}
\end{equation}
At isostaticity, these bounds are saturated, leading to the value for the isostatic coordination number
\begin{equation}
\langle z_g \rangle_{\rm iso} = 4 - \frac{2}{N_G}.
\label{z_iso_equation}
\end{equation}

Another derivation of the above isostatic condition can be arrived at 
by considering the triangulation graph of the packings.
The fictitious contacts provide the missing equations needed to determine the positions of the grains, and in turn the contacts, completely.
This can be parametrized in terms of the angles of the triangulation, given the   position of a single disk, the angles of the triangulation completely determine the positions. In this case the undetermined quantities are the ``ficititious angles" that are formed between the grain distance and fictitious vectors, and between two fictitious vectors. Using the local condition for the number of triangles and the fact that the number of determined contact angles for each void is $z_v$, we find the undetermined angles belonging to each void is equal to $2 z_v - 6$.
Summing over the entire system, and this time accounting for a global translational degree of freedom leads to
\begin{equation}
N_{\rm dof} = 2 N_C - 6 N_V - 2.
\end{equation}
All the constraints in the system can then be derived from the constraints on each of the basic triangular units. For example, every loop constraint can be obtained by summing the triangle constraints in the interior of the loop. Therefore
\begin{equation}
N_{\rm constraints} = N_T.
\end{equation}
We can then use the number of triangles from Eq. (\ref{triangles1}),
and the criterion $N_{\rm dof} \ge N_{\rm constraints}$ to obtain the bound on the number of undetermined angles, leading directly to Eq. (\ref{isostatic_inequality}).

\section{Distribution Functions}
\label{distribution_functions_section}

We next perform simulations in order to generate contact networks of jammed frictionless disks close to the transition. We then use these configurations to study distribution functions of the underlying geometrical degrees of freedom such as contact angles, lengths of triangulation vectors, and the areas of the grain and void polygons.

\subsection{Simulation Protocol}
\label{protocol_section}

As the protocol dependence of jamming is well-known \cite{majmudar_behringer_prl_2007,onada_liniger_prl_1990,
bertrand_pre_2016,berthier_sollich_prl_2011,
nowak_jaeger_pre_1998,bi_nature_2011} care must be taken in producing configurations.
The configurations we study are produced using a variant of the O'Hern protocol that effectively samples all available energy minima in an unbiased manner \cite{ohern_prl_2002,ohern_pre_2003}.
We use periodic boundary conditions in both directions, the configurations are in a box size $1$ (i.e. $L_x=L_y=1$). 
The protocol to generate a packing is as follows:\\ 
(1) We randomly place grains (circular disks) in the box,\\ 
(2) We then minimize the potential energy in Eq. (\ref{full_system_energy}) using non-linear conjugate-gradient,\\
(3) Once an energy minimum to within a desired tolerance ($10^{-16}$ in our case) is reached, we change the grain size.\\ 
Steps (2) and (3) are repeated until a packing is found. 
The grain size is changed as follows: after energy minimization the configuration has an energy per particle $E_G$. The grains are grown if $E_G < E_{\rm min}$, and shrunk if $E_G > E_{\rm max}$, where $E_{\rm min}$ and $E_{\rm max}$ represent a tolerance window for our marginally jammed configurations.
We shrink or grow the particles according to a parameter $\Delta > 0$ as
\begin{equation}
\sigma_g = \sigma_g/(1+ \Delta) ~~{\rm or}~~  \sigma_g (1+ \Delta).
\end{equation}
The disks are grown or shrunk until the un-minimized energy of the packing falls outside the tolerance window.
We reduce the step size $\Delta$ by half after every step of energy minimization. 
Our marginally jammed configurations have energies per particle between $1 \time 10^{-16}$ and $1.1 \times 10^{-16}$.

\subsection{Energy Ensemble}
\label{energy_ensemble_section}

An important open question in the field of granular materials is that of ensembles. In the usual statistical mechanics framework one takes a simple weighted sum of all available states of the system, with a weight chosen using an energy functional and the temperature. In the case of granular materials, the system is inherrently ``athermal", i.e. there is a very weak dependence of the properties of the system with temperature. The natural question then becomes, how does one group the configurations for an athermal system? There are several proposed methods of generalizing ensembles to granular systems including grouping by the total stress, the configurational entropy and also using real space volumes \cite{henkes_chakraborty_ohern_prl_2007,edwards_physicaa_1989,frenkel_molphys_2013,blumenfeld_prl_2015}.

For soft disks if one focusses solely on the ensemble of jammed states, as we are doing, the energy of the system represents a continuously tunable parameter close to the transition. In this sense we approach the transition only from stressed states with well defined energies and all states below the transition are absent from our ensemble.
There is evidence that grouping configurations by the energy of configurations does indeed yield accurate statistical results \cite{henkes_chakraborty_pre_2009, henkes_chakraborty_ohern_prl_2007}.
We define configurational averages of system properties in the {\it microcanonical ensemble} as
\begin{equation}
\langle x (E_G)\rangle = \frac{\sum_{\Omega} \delta(E(\Omega) - E_G) x(\Omega)}{\sum_{\Omega}  \delta(E(\Omega) - E_G)},
\label{configurational_average}
\end{equation}
where $\Omega$ refers to an individual configuration, $x(\Omega)$ represents the value of a system property for the configuration, $E(\Omega)$ refers to the energy of the configuration and the summation is taken over all configurations $\Omega$. In practice, since obtaining statistics for large system sizes is computationally intensive, we allow for a finite energy width in our samples of $[E_G, 2 E_G]$. We have tested that this does not significantly change the statistics.
Similarly we can define distribution functions 
\begin{equation}
P_{E_G}(x) = \langle \delta (x - x(\Omega)) \rangle,
\label{distribution_definition}
\end{equation}
where the angular brackets represent configurational averages as in Eq. (\ref{configurational_average}). The moments of system quantities can then be computed as
\begin{equation}
\langle x^n (E_G)\rangle = \int x^n P_{E_G}(x) dx.
\label{moment_definition}
\end{equation}

We perform simulations on bidispersed systems with a ratio of grain diameters  $1:1.4$ (referred to as type-$A$ and $B$ respectively). We simulate systems with  an equal number of disks of type-$A$ and type-$B$.
 We use bidispersed disks to avoid crystallization as we are interested in the disordered structures that arise near the transition. The diameter ratio $1:1.4$ ensures incommensurability within numerical error, upto the sizes that we measure and has been well studied in the literature \cite{ohern_prl_2002}.
We measure various properties of the networks formed by the disks in order to characterise their spatial structure near the jamming transition.

\subsection{Self Averaging Quantities}
\label{self_averaging_section}
The structural properties of jammed systems displays randomness that arises from the constraint satisfaction of a large number of degrees of freedom (Eq. (\ref{force_balance_equation})). This is very similar to properties of structural randomness arising from quenched disorder in glassy systems \cite{parisi_zamponi_rmp_2010}.
In order to assign microscopic degrees of freedom to the jammed packings it becomes relevant to test which quantities in the system display self averaging properties, i.e. 
any physical property $x$ of the system, such that 
\begin{equation}
\frac{\langle x^n \rangle_\Omega}{\langle x^n \rangle} \to 1 ~~{\rm as}~~ N_{\Omega} \to \infty,
\end{equation}
where $\langle \rangle_{\Omega}$ represents an average over a single configuration $\Omega$ of size $N_{\Omega}$ and $\langle \rangle$ represents an average over all configurations. In practice, averages up to the second moment are good enough to test this behaviour as the central limit theorem becomes valid. Such an ensemble can then be completely described by a single large system, and one can define reproducible distribution functions as in Eq. (\ref{distribution_definition}). Such quantities can then be used to construct extensive variables that can be described by microscopic theories. We find that the distributions of contact angles, lengths of contact vectors, the areas of grain and void polygons and consequently the total covered area, displays self averaging behaviour. This is indicative of only short-range order close to the unjamming transition.
Self averaging behaviour can be destroyed by long range correlations, and therefore it is necessary to test under what circumstances such behaviour is valid. In section \ref{correlation_section} we test the length scales up to which real space correlations persist in such packings.

\begin{figure}
\hspace{0.5cm}
\includegraphics[height=.35\columnwidth,angle=0]{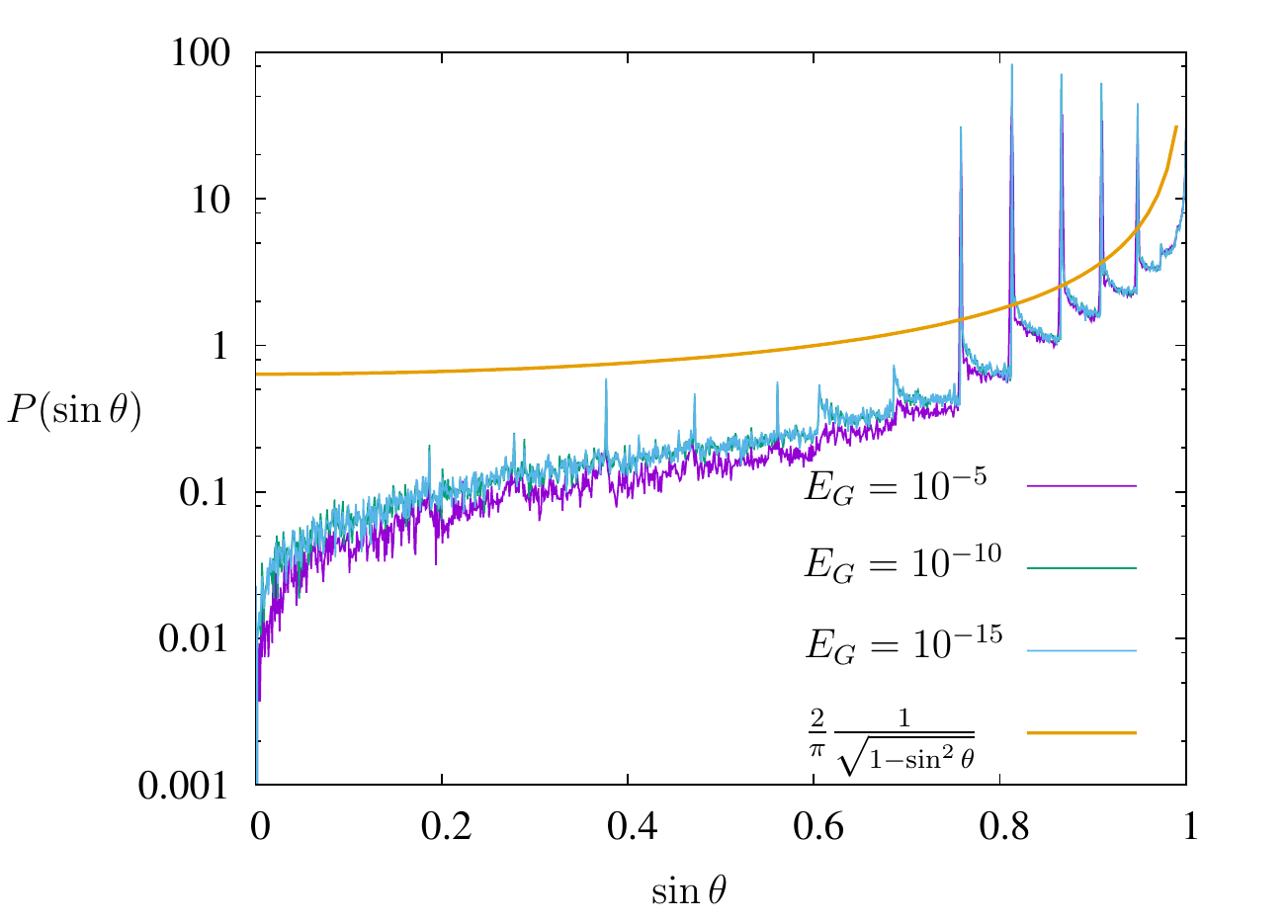}
\includegraphics[height=.35\columnwidth,angle=0]{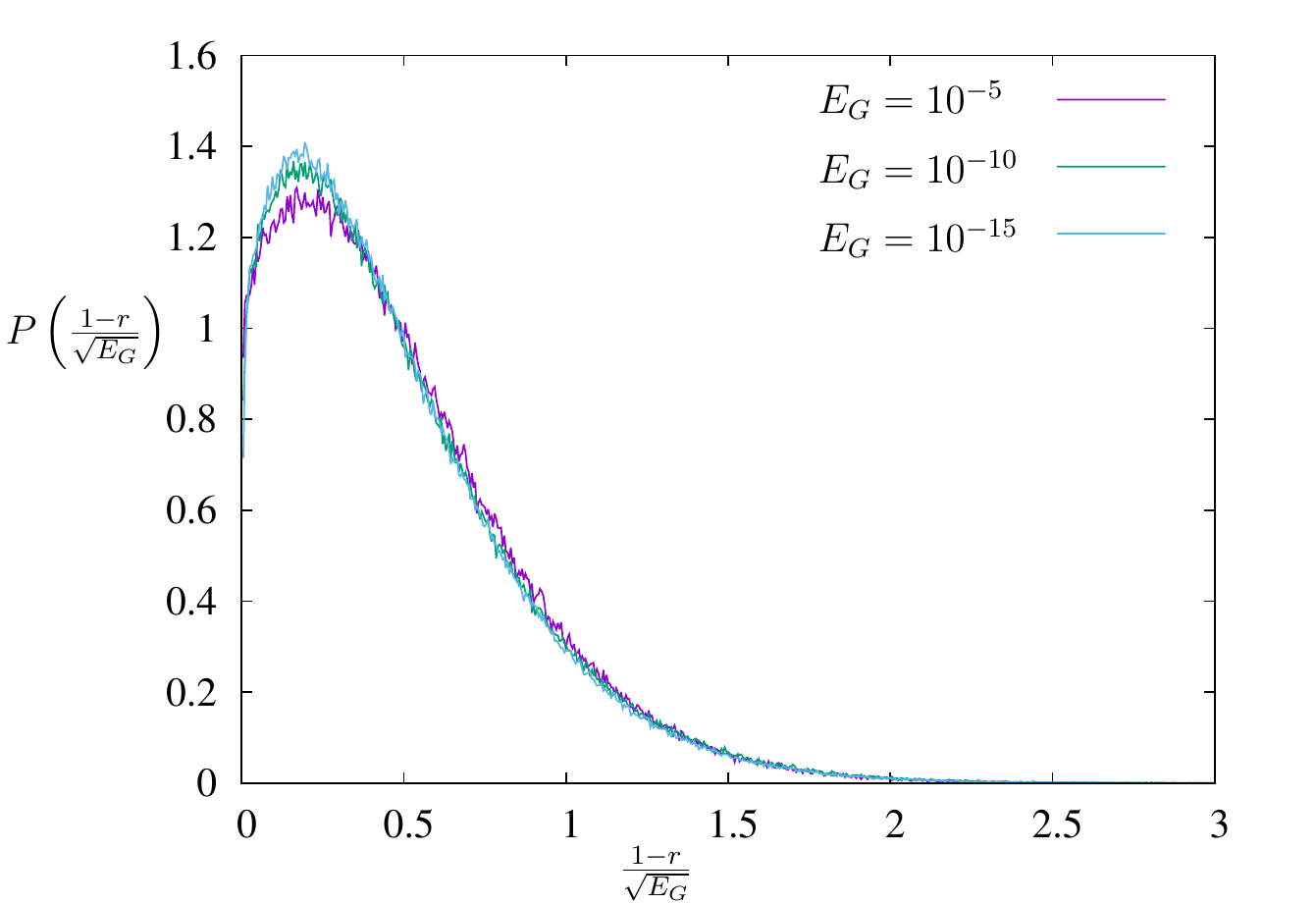}
\caption{{\bf (Left)} Distribution of relative contact angles $\theta = \theta_{g,c'} - \theta_{g,c}$. These relative contact angles display random behaviour except at well-defined points that corresponds to ordered structures formed by three of more disks in contact. The five most prominent peaks can be identified as arising from three disks in contact. {\bf (Right)} Distribution of lengths of contact vectors measured in packings with $N_G = 2048$ at different energies.  The plot shows the distribution of $(1 - r)/\sqrt{E_G}$, where $r = |\vec{r}_{g,c}|/\sigma_g$. This distribution displays the well-studied force distribution curve \cite{wyart_prl_2012}.
}
\label{prob_angle_figure}
\end{figure} 

\subsection{Contact Vectors and Contact Angles}
\label{contact_vectors_section}

We first investigate the distribution of contact angles that form in the packings of bidispersed disks close to the unjamming transition. We measure the distribution of relative angles $\theta = \theta_{g,c'} - \theta_{g,c}$, where $c$ and $c'$ are chosen cyclically within each grain.
This quantity has been well-studied in the literature \cite{silbert_grest_landry_pre_2002}. In Fig. \ref{prob_angle_figure} we plot the distribution $P(\sin \theta)$, measured in packings of $N_G = 2048$ disks at different energies. Also plotted alongside is a distribution of completely random relative angles $P(\sin \theta) = \frac{2}{\pi} \frac{1}{\sqrt{1 - \sin^2 \theta}}$.
We find that the relative contact angles display a random behaviour except at well-defined points that we can identify as ordered structures. We discuss these structures, and specifically those formed by three disks in contact in detail in Section \ref{measures_of_order_section}. For the diameter ratio $1:1.4$, the peaks can be shown to occur at $\theta = 1.24565, 1.14102, 1.0472, 0.94797$ and $0.859551$.

We next measure the distribution of the lengths of the contact vectors formed in the packings. In order to account for the changes in grain radii between different configurations we normalize the length of the contact vectors by the radius of the grain to which they belong as $r = |\vec{r}_{g,c}|/\sigma_g$.
In the case of linear spring potentials as we are studying, this distribution is equivalent to measuring the distribution of forces in the system. This can be seen by taking a first derivative of Eq. (\ref{potential_energy_functional}) with respect to the distance vector $\vec{r}_{g,g'}$. Since the sum of squares of these individual forces is equal to the total energy in the system it is natural to normalize the contact vector lengths by $\sqrt{E_G}$. In Fig. \ref{prob_angle_figure} we plot the distribution of $(1- r)/\sqrt{E_G}$. We find that the lengths of contact vectors scaled with this factor displays the well-studied force distribution curve. This distribution increases as a power law at small lengths $P(x) \sim x^{0.17}$ as $x \to 0$ similar to those observed in \cite{wyart_prl_2012}, and falls with the well-known exponential decay $P(x) \sim \exp(-x)$ as $x \to \infty$.

\subsection{Connectivity}
\label{connectivity_section}
\begin{figure}
\hspace{0.5cm}
\includegraphics[height=.35\columnwidth,angle=0]{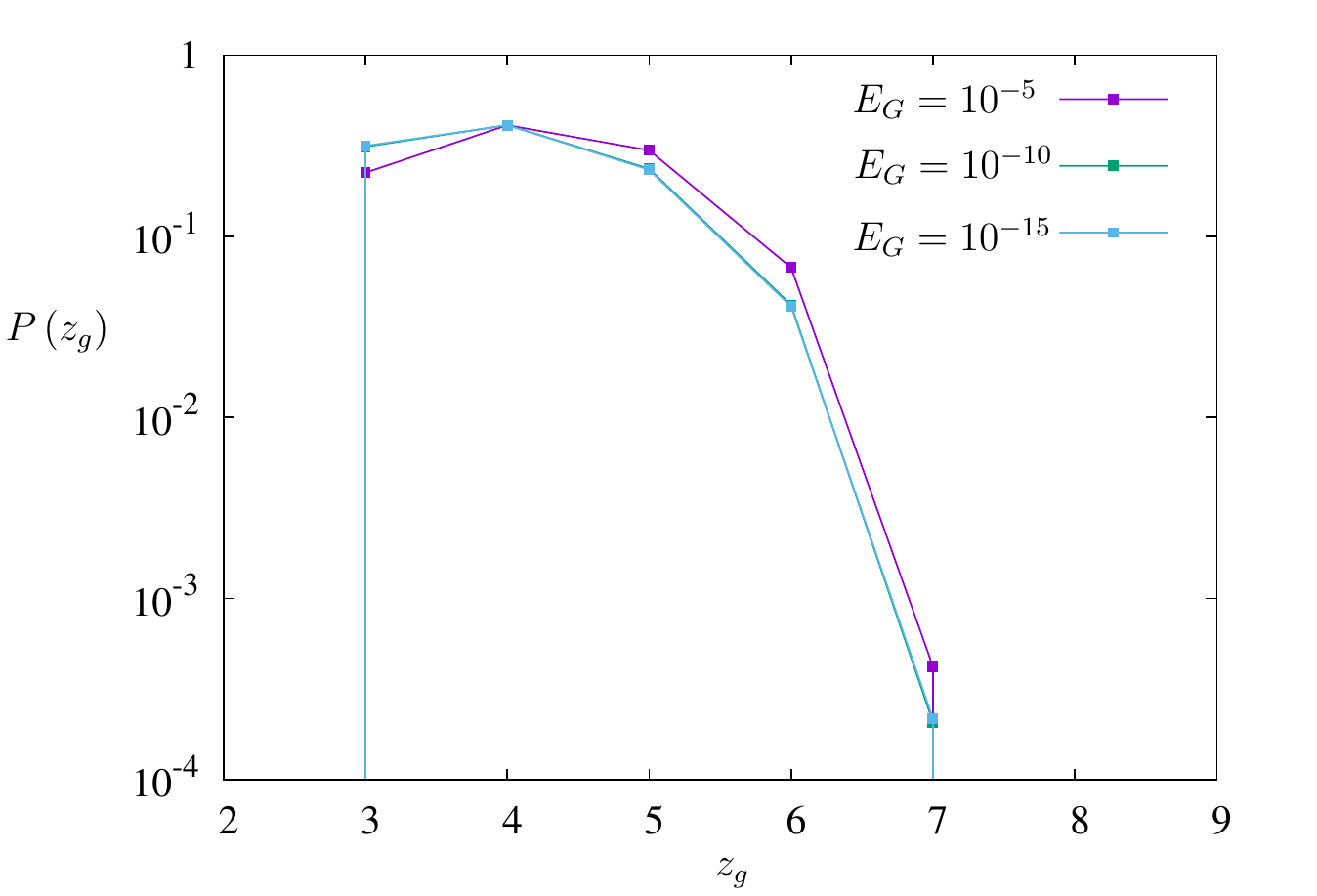}
\includegraphics[height=.35\columnwidth,angle=0]{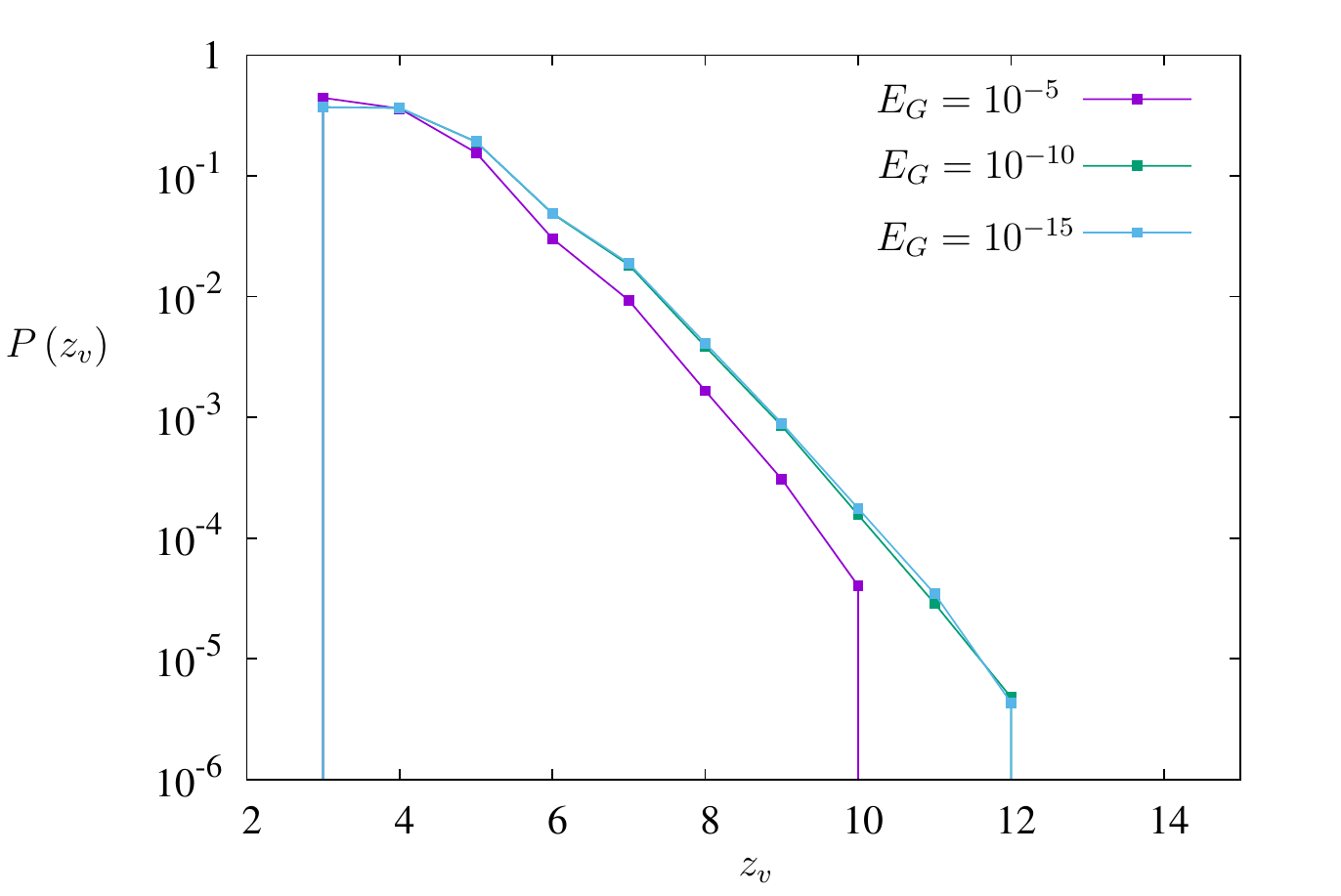}
\caption{{\bf (Left)} Distribution of the connectivities of the grains. The overlap constraints limit $z_g \le 7$ for the ratio of grain diameters $1:1.4$ and the energies that we consider.
{\bf (Right)} Distribution of the connectivities of the voids. The connectivities of the voids are unconstrained, however we find an exponential behaviour for large $z_v$ as $E_G \to 0$. Since the connectivity graphs for grains and voids are dual to each other, their average connectivities are related by the duality relation $\frac{1}{\langle z_g \rangle} + \frac{1}{\langle z_v \rangle} = \frac{1}{2}$.}
\label{z_figure}
\end{figure} 

The next question we address is the connectivities of the grain and void polygons that we investigated in Section \ref{coordination_section}.
Although the isostaticity condition predicts the average values of the grain and void connectivities, the question of their distributions is non-trivial.
It has been noted in several studies that $\langle z_g \rangle$ attains its exact isostatic value for marginal states. In addition, recent studies have found that individual sections of marginal states are {\it locally isostatic} \cite{ellenbroek_prl_2015}. This suggests that one can define connectivities of the network at the local level. 
To test this we measure the distributions of 
connectivities of the grains and also of the voids near the unjamming transition.
We find that these distributions do indeed display self averaging behaviour.
Since the connectivity graphs for grains and voids are dual to each other, their average connectivities are related by the duality relation $\frac{1}{\langle z_g \rangle} + \frac{1}{\langle z_v \rangle} = 2$ (Eq. \ref{duality_relation}). 
Close to the unjamming transition the average connectivities approach their isostatic values $\langle z_g \rangle = \langle z_v\rangle \to 4$. We plot these distribution functions in Fig. \ref{z_figure}.
The connectivity of each grain is limited by geometrical constraints. Using the allowed angles between contact vectors, the maximum allowed connectivity of a grain can be computed as $z_g \le \lfloor 2 \pi/\theta_{\rm min} \rfloor = 7$. In contrast, the connectivities of the voids are unconstrained. We find that the distribution of void connectivities exhibits an exponentially decaying behaviour as $E_G \to 0^+$ (see Fig. \ref{z_figure}).

\section{Grain Area}
\label{grain_area_section}
In this section we focus on the total area occupied by the grain polygons. We measure statistical properties of these areas and provide evidence that this polygonal measure can be used as a reliable jamming parameter in a manner similar to the packing fraction.

\begin{figure}
\hspace{0.5cm}
\includegraphics[height=.35\columnwidth,angle=0]{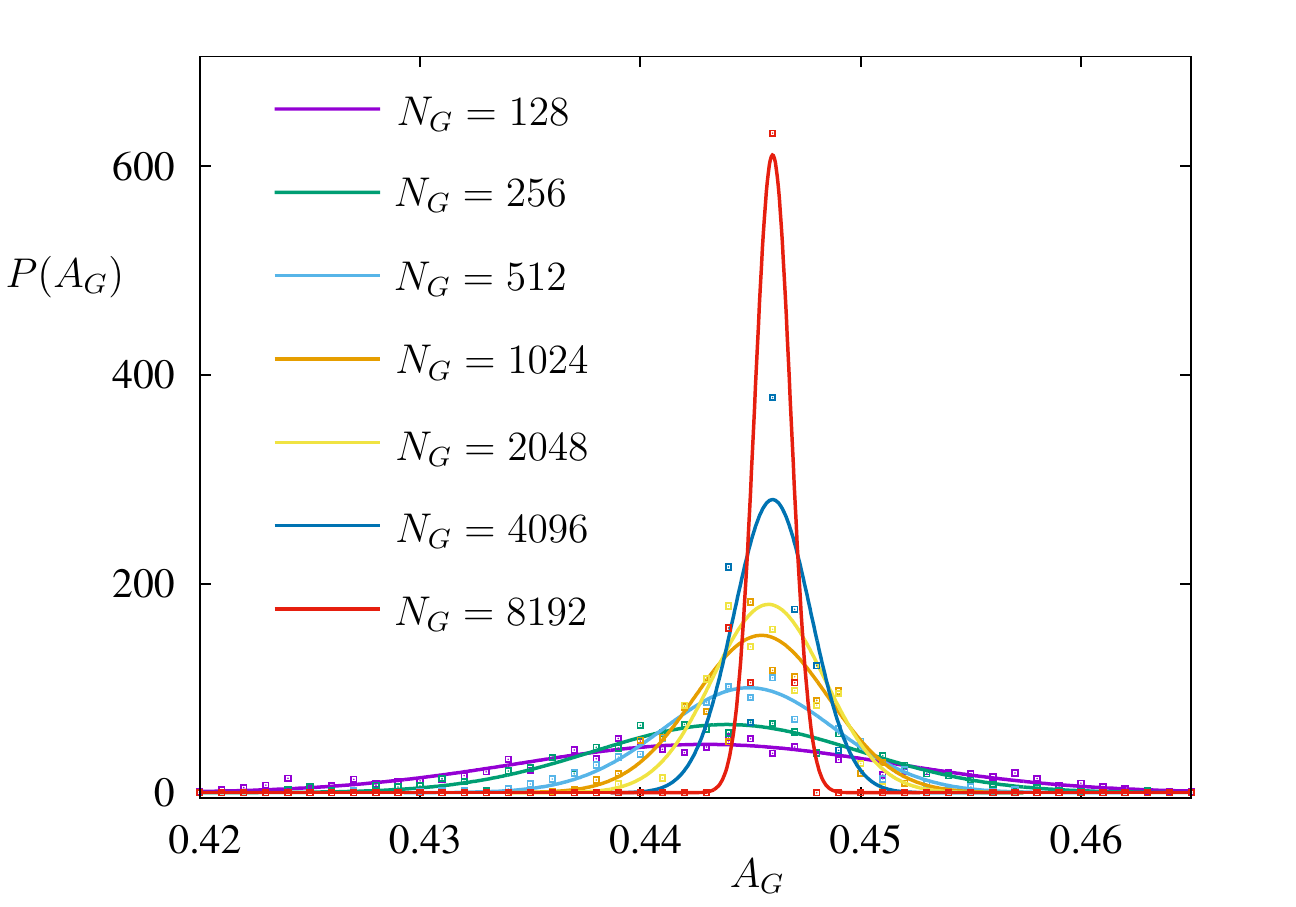}
\includegraphics[height=.35\columnwidth,angle=0]{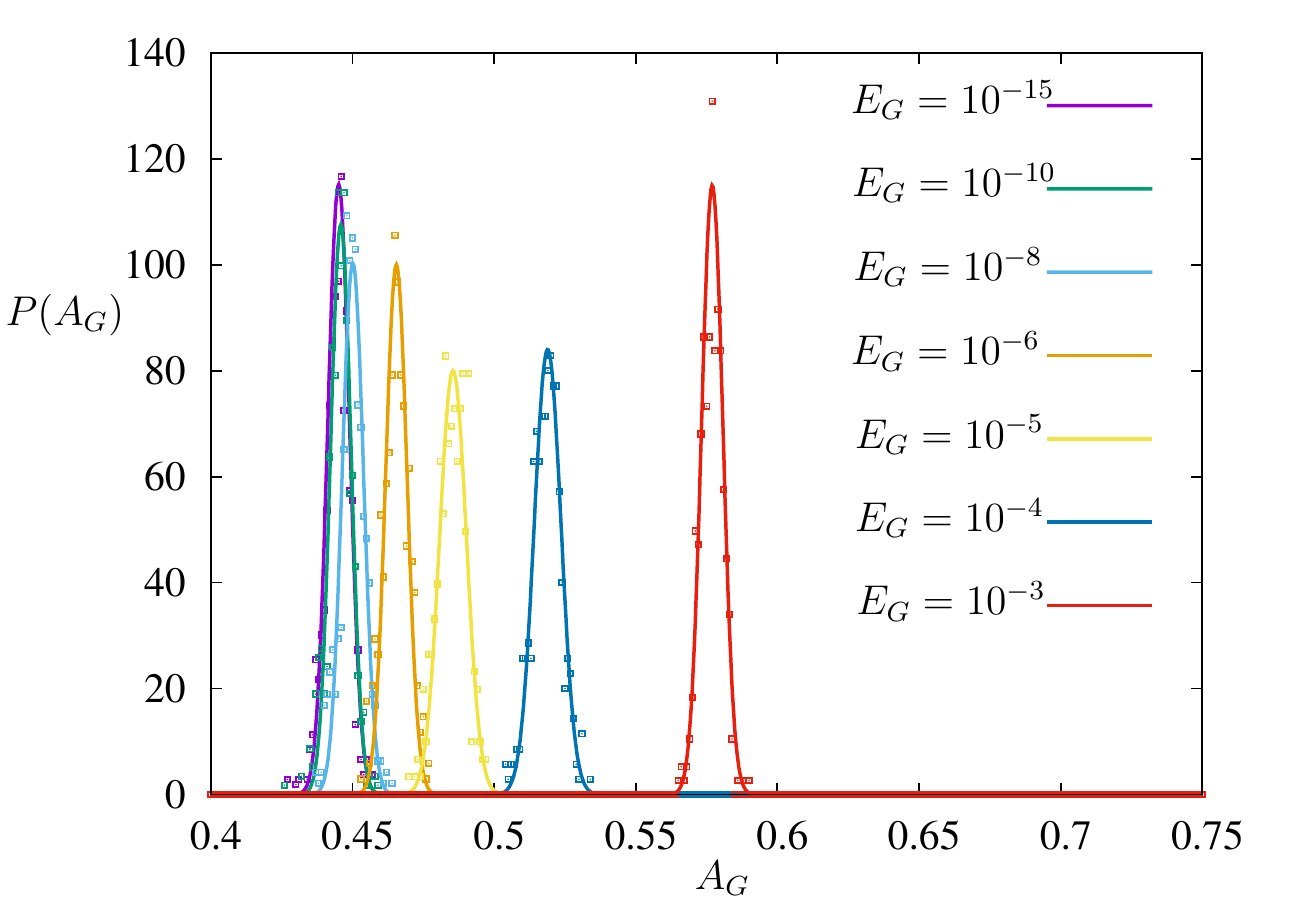}
\caption{{\bf (Left)} Distribution of the total area covered by the grain polygons $A_G$ in the packings of bidisperse grains at  the lowest measured energy $E_G = 10^{-15}$. The continuous curves represent the best-fit Gaussians associated with the distributions. As the number of grains is increased the width of the distribution decreases. Using finite-size scaling fits we find is $A_{G}^* = 0.446(1)$ as $N_G \to \infty$ and $E_G \to 0^{+}$ (see Table \ref{finite_size_scaling_table}). {\bf (Right)} Behaviour of the grain area distributions for different energies $E_G = 10^{-15}$ to $10^{-3}$ for packings of $N_G = 512$ disks. We find that the grain areas serve as reliable parameters with which to measure system properties at these energies.}
\label{area_figure}
\end{figure} 

\subsection{Grain Area as a Reliable Jamming Parameter}
\label{reliable_parameter_section} 

One of the fundamental problems in the field of granular materials is the reliable definition of packing fractions as a measure of occupied volumes or areas. The standard way of computing packing fractions for soft particles $\phi = \sum_{g} \pi \sigma_g^2$ involves the summation of volumes of undistorted particles. At high energies this overestimates the true covered fraction of area as overlaps are also included. Another deficiency of standard packing fractions is the treatment of rattlers that are not part of the contact network and therefore do not contribute to the stability of the network.
In order to get reliable estimates of exponents from packing fractions, the areas of rattlers need to be included in $\phi$. Such rattlers are naturally absent from the polygonal measure. Another issue with packing fractions becomes clear when one tries to probe {\it local} properties. It becomes harder to define such properties at the local level. Although Voronoi volumes have been considered in the literature for a long time, they fail to provide information about important quantities such as porosity \cite{donev_stillinger_prl_2005}. The grain area differs crucially from the measure of Voronoi volumes as it allows us to measure the partitioning of space into grain and void sections, thereby allowing us to probe the structural properties that cause jamming. If one aims to construct a density based order parameter for the transition, the  following properties should be satisfied: the sum of local order parameters should provide the global order parameter and it should scale with system size. It is easy to see that the first non-trivial parameter that one can construct using the local variables that satisfies the above properties is the grain area.
In our case $A_G$ also serves as the area-fraction since the total space area is $1$ ($L_x = L_y = 1$).

A primary quantity of interest in the polygonal representation is then the total area occupied by the grain polygons in jammed packings and in particular, near the unjamming transition. The total fraction of area occupied by the circular areas of disks at the transition is well known to be $\phi_J \approx 0.84$ \cite{ohern_prl_2002}, which is relatively close to the hexagonal ordered structure $\phi = \pi/\sqrt{12} \approx 0.9069$. In contrast, we find that the ``reduced" area defined by the polygonal construction for packings near the transition differs significantly from the reduced area of the close packed structure.
In order to characterise these grain areas we measure $P(A_G)$, the probability of finding a packing with total covered area $A_G$ in the ensemble. We measure this distribution for packings created with different numbers of grains at varying energies near the unjamming transition. In Fig. \ref{area_figure} we plot the measured distributions of $P(A_G)$ for systems with increasing densities of particles at the lowest measured energy. We find that this probability distribution does indeed get sharper as the density of grains is increased. We find that for the highest densities we have measured, the width of these distributions scales as $1/\sqrt{N_G}$, indicating that we can attribute local areas to such packings. We also find that this area varies in a well defined manner with energy (plotted in Fig. \ref{area_figure}), allowing us to use it as a parameter to study the transition as $E_G \to 0^{+}$, and also at higher energies.

\begin{table}
\hspace{3cm}
    \begin{tabular}{| l || l | l || l | l |}
    \hline
     $N_G$ & $A_G^*(E_G = 0)$ & $\beta_E$ & $A_G^*(\Delta Z = 0)$ & $\beta_Z$\\ \hline \hline
     $128$ & $0.4416(2)$ & $0.299(3)$& $0.4424(2)$ &$1.02(2)$\\ \hline
     $256$ & $0.4423(2)$ & $0.286(3)$& $0.4431(2)$ &$1.01(2)$\\ \hline
	 $512$ & $0.4446(2)$ & $0.285(3)$ & $0.4448(2)$ & $1.001(2)$\\ \hline
     $1024$ & $0.4454(2)$ & $0.284(3)$ & $0.4453(2)$ & $1.001(2)$\\ \hline
     $2048$ & $0.4458(2)$ & $0.284(3)$ & $0.4458(2)$ & $1.001(2)$\\ \hline
     $4096$ & $0.4461(2)$ & $0.284(3)$ & $0.4460(2)$ & $1.001(2)$\\ \hline
     $8192$ & $0.4461(4)$ & $0.284(4)$ & $0.4460(3)$ & $1.001(3)$\\ \hline
    \hline
  	\end{tabular}
  	\caption{Finite size scaling estimates for the transition point and scaling exponents using systems with varying number of disks $N_G$. We can estimate the unjamming point ($A_G^*$) from both $E_G \to 0^+$ (left), and $\Delta Z = \langle z_g \rangle - \langle z_g \rangle_{\rm iso} \to 0^+$ (right). Using finite size scaling and accounting for statistical sampling error, we find $A_G^* = 0.446(1)$, $\beta_E = 0.28(2)$ and $\beta_Z = 1.00(1)$.}
\label{finite_size_scaling_table}
\end{table}

\subsection{Critical Exponents and Finite Size Scaling}
We next use the total grain area to study critical exponents of the unjamming transition.
We test the scaling behaviour of the total grain area with two parameters  that acquire well defined values at the unjamming transition, namely the energy per grain $E_G \to 0^+$ and the excess coordination $\Delta Z = \langle z_g \rangle - \langle z_g \rangle_{\rm iso} \to 0^+$. The unjamming transition is located precisely at $E_G = 0$, and at $\Delta Z = 0$. The value of $\langle z_g \rangle_{\rm iso}$ was derived in Eq. (\ref{z_iso_equation}).
 We can therefore use the deviations from these two values to test the scaling of the total area $A_G$.
Since the polygonal measure is only sensitive to structural changes in the configurations (a simple expansion or contraction of all disks is given the same weight), it is interesting to see whether we can measure exponents that provide information about the geometric structure of the packings.
We find that this is indeed the case, and that critical exponents measured from grain areas differ from those measured using standard packing fractions.

As the unjamming transition is approached from above $E_G \to 0^{+}$, fewer disks are in contact, leading to a decrease in the total grain area. We plot the distributions of the area for different energies for $N_G = 512$ disks in Fig. \ref{area_figure}.
For higher densities, for example with $N_G = 8192$,
we find $A_G \approx 0.482$ for $E_G = 10^{-5}$, $A_G \approx 0.448$ for $E_G = 10^{-10}$ and $A_G \approx 0.446$ for $E_G = 10^{-15}$. It is then interesting to ask at what covered fraction of total area the transition occurs. In order to estimate this quantity, as is standard in the study of phase transitions, we use finite-size extrapolation. We first estimate the transition point for a given system size (number of disks) as $E_G \to 0^+$, and then extrapolate to $N_G \to \infty$. 
We can estimate these critical points using both $E_G$ and $\Delta Z$. The locations of these points for different system sizes is summarized in Table \ref{finite_size_scaling_table}. We find that the finite size effects scale as $1/N_G$ with the system size. Using these values, and accounting for statistical sampling error, we find 
\begin{equation}
A_G \to A_G^* = 0.446(1) ~~\textmd{as}~~ E_G \to 0^+  ~~\textmd{and}~~ N_G \to \infty.
\end{equation}
$A_G^*$ therefore defines a new real-space parameter for the unjamming transition. We note that the polygonal construction produces $A_G = 3/4$ for the hexagonal ordered structure, with the value $0.446$ significantly different from this number as compared to the closeness of the values in the packing fraction measure.

\begin{figure}
\hspace{0.5cm}
\includegraphics[height=.35\columnwidth,angle=0]{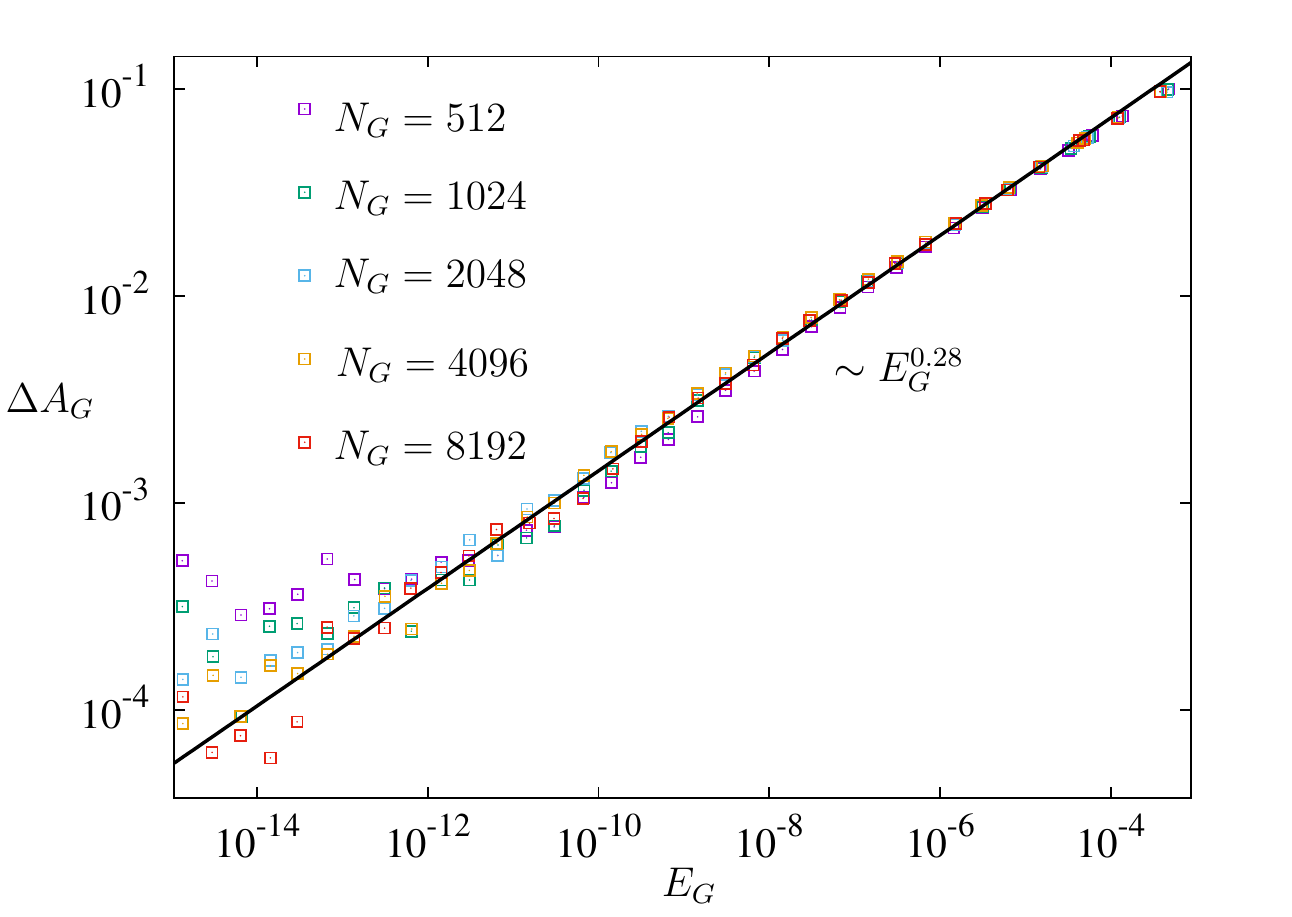}
\includegraphics[height=.35\columnwidth,angle=0]{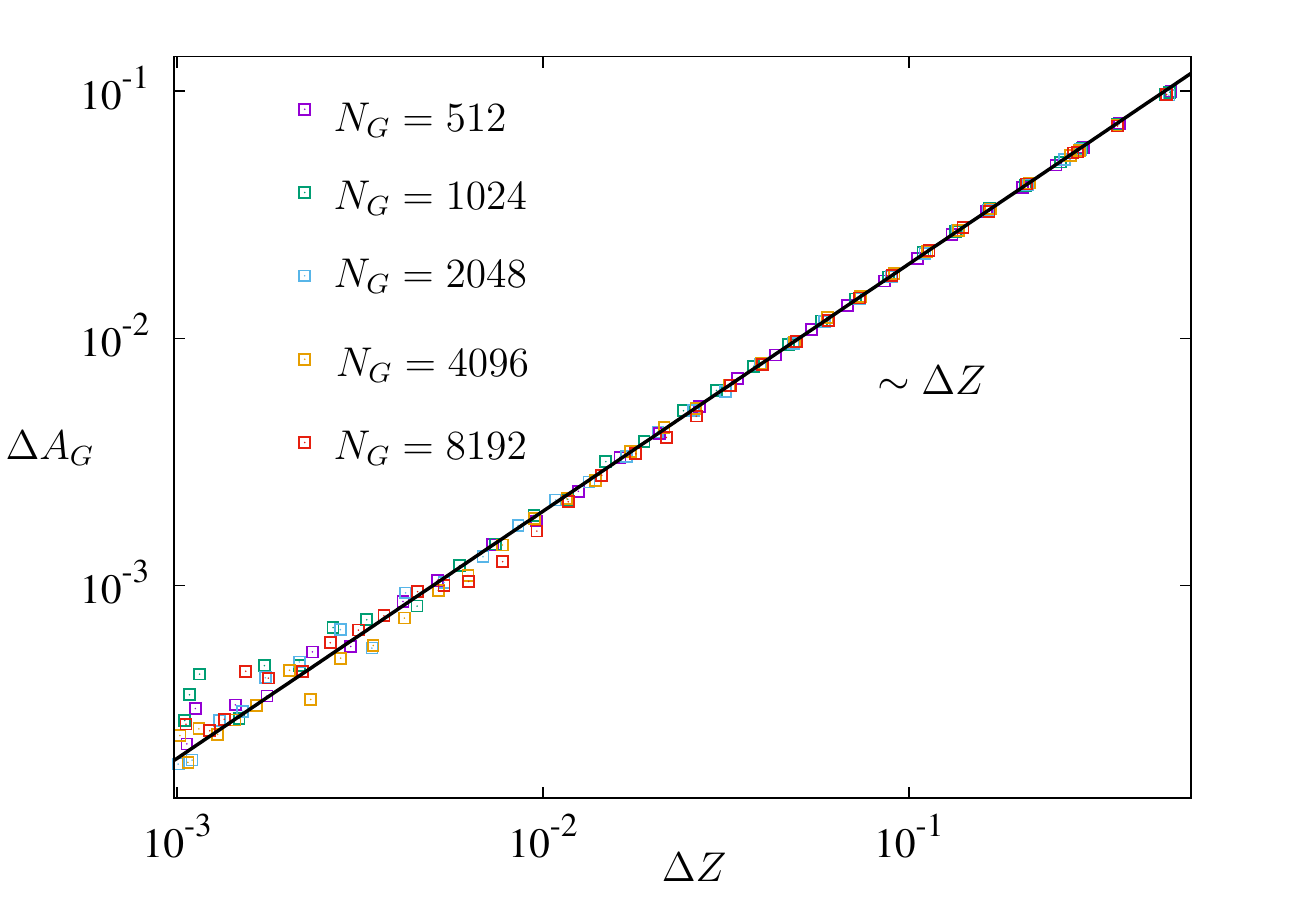}
\caption{{\bf (Left)} Scaling of the excess grain area $\Delta A_G = A_G - A_G^*$ with total energy per particle $E_G$. We find that the excess grain area scales as a power of the total energy in the system with exponent $\beta_E = 0.28(2)$. {\bf (Right)} Scaling of $\Delta A_G$ with excess coordination in the system $\Delta Z$. We find that the excess grain area scales as a power of $\Delta Z$ with exponent $\beta_Z = 1.00(1)$.}
\label{scaling_with_EG_figure}
\end{figure} 

We next use the deviation of the total grain area from its value at the unjamming point $\Delta A_G = A_G - A_G^*$ to test the scaling behaviour with our two control parameters $E_G$ and $\Delta Z$.
We define two scaling exponents $\beta_E$ and $\beta_Z$ that quantify this behaviour
\begin{eqnarray}
\nonumber
\Delta A_G \sim (E_G)^{\beta_E},\\
\Delta A_G \sim (\Delta Z)^{\beta_Z}.
\end{eqnarray}
In Fig. \ref{scaling_with_EG_figure} we plot
the scaling of $\Delta A_G$ with total energy per particle $E_G$. 
We find that the excess grain area scales as a power of the total energy in the system with exponent $\beta_E = 0.28(2)$. The estimates of $\beta_E$ from finite size extrapolation are summarized in Table \ref{finite_size_scaling_table}. This value of $\beta_E$ is in contrast with the known scaling of the packing fraction which displays an exponent $\approx 0.5$ \cite{ohern_prl_2002}. This is a surprising feature of the polygonal measure, in that even though it measures the occupied area similar to the packing fraction, it has different scaling properties close to the transition. We next test the scaling of the total grain area with excess coordination near the transition.
In Fig. \ref{scaling_with_EG_figure} we plot the scaling of $\Delta A_G$ with $\Delta Z$. We find that the excess grain area scales as a power of $\Delta Z$ with exponent $\beta_{Z} = 1.00(1)$. Once again this is in contrast with the scaling of packing fractions which displays an exponent $\approx 2.0$ \cite{ohern_prl_2002}.  Estimates of $\beta_Z$ from finite size extrapolation are summarized in Table \ref{finite_size_scaling_table}.
The linear scaling of $\Delta A_G$ with $\Delta Z$ suggests an assignment of individual areas to the contacts which can serve as a microscopic order parameter. We find that the edge triangles defined in Section \ref{Polygonal_representation_section} serve as precisely such variables. We measure the statistics of these edge triangles and other geometrical characteristics of the underlying packings in Section \ref{measures_of_order_section}.

\section{Measures of Order}
\label{measures_of_order_section}
In this section we characterize the types of polygonal tilings that
emerge in the packing of disks near the marginally jammed state. 
The area statistics of the individual grain polygons and the void polygons provide useful insight into the structural randomness present at the microscopic scale.

\subsection{Edge Triangles}
\label{edge_triangles_section}
One quantity that can be used to measure local disorder is the shape of the grain polygons formed by the jammed packings. 
In order to characterize the shape of these grain polygons, we measure the areas of their constituent edge triangles introduced in Section \ref{Polygonal_representation_section}. 
At exactly $E_G = 0$, the distribution of areas of edge triangles is directly related to the distribution of contact angles. At finite energies these differ as the edge triangles also includes the effect of overlaps between grains.
The areas of the individual triangles can vary between $0$ and $\frac{1}{2} {\sigma_g}^2$ where $\sigma_g$ is the radius of the grain to which they belong (see Fig. \ref{system_figure}). We note that $\{ \sigma_g \}$ can vary between different configurations chosen with the same energy $E_G$. In order to account for the different sizes between configurations, we measure the following normalized area $\alpha_{e} = a_{g,e}/{\sigma_g}^2$. This can take values between $0$ and $\frac{1}{2}$. The minimum value is not attained in ordinary packings at low compressions, however we can obtain good statistics of these triangular areas close to the maximum, corresponding to disks with contact angles close to $\theta_{g,c'} - \theta_{g,c} = \pi/2$.
 We plot the distribution of edge triangle areas $P(\alpha_e)$ for a system with $N_G = 2048$ disks at different energies in Fig. \ref{grain_void_figure}. 
We find that this distribution displays self averaging behaviour and reaches a limiting form as the system size is increased. This is significant since it is then possible to distribute the total grain area $A_G$ into the ``local" areas of edge triangles. The edge triangles therefore serve as reliable local jamming parameters that can be used to construct microscopic theories.

\begin{figure}
\hspace{1cm}
\includegraphics[width=1\columnwidth,angle=0]{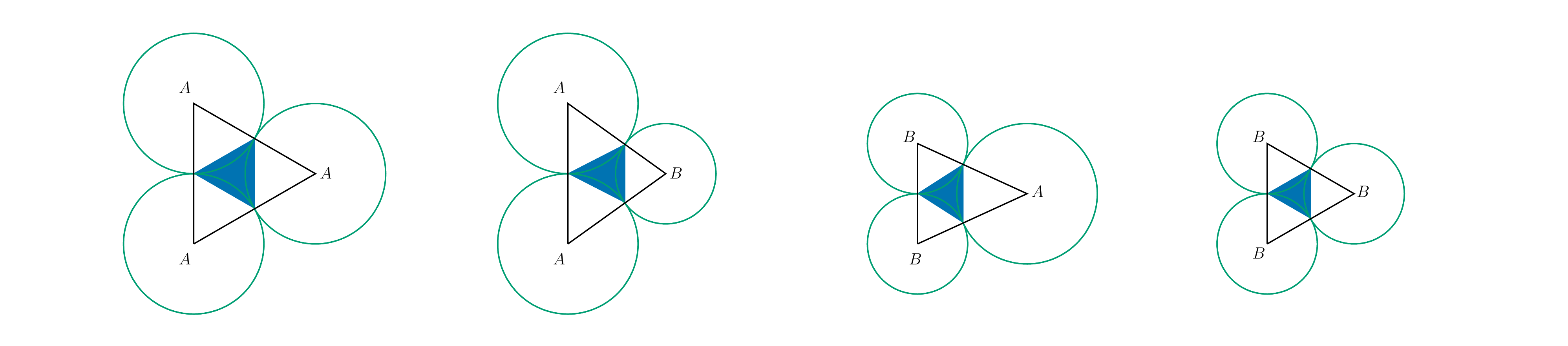}
\caption{The four ordered structures arising from three disks in contact that can occur in bidispersed systems. This causes ordered peaks to appear in the distribution of edge triangle areas (white triangles). For the diameter ratio $1:1.4$ these peaks occur at $\alpha_e = a_{g,e}/\sigma_g^2 = 0.473803, 0.45453, 0.433013, 0.406116$ and $0.378775$ (See Fig. \ref{grain_void_figure}). These structures also cause peaks to appear in the areas of void polygons (blue triangles). For the diameter ratio $1:1.4$ these peaks occur at $\alpha_v = a_{v}/{\sigma_A}^2 = 0.433013, 0.33843, 0.270553$, and $0.220925$ (See Fig. \ref{grain_void_figure}).
The fraction of the system in these ordered structures gives us a measure of order in the system $\Psi_O$.}
\label{three_disks_figure}
\end{figure} 

We next focus on the well-defined peaks in the distribution of edge triangle areas. We find that these peaks become sharper as the transition is approached. These peaks can be identified as arising from ordered structures formed by three, four and higher numbers of disks in contact.
The first five peaks in the distribution of $\alpha_e$ can be identified as arising from three disks in contact. These arise from combinations of type-$A$ and type-$B$ disks, namely $A(AA)\equiv B(BB)$, $A(AB)$, $B(AB)$, $A(BB)$ and $B(AA)$, where the brackets represent disks in contact with the unbracketed disk, to which the edge belongs.
At the  marginally jammed state (with zero overlap), the distances between the disks in contact can take only unique values (see Fig. \ref{three_disks_figure}) leading to unique values of the angles and consequently the areas which can be computed exactly. For the ratio of diameters $1:1.4$ these are $\alpha_e^{A(AA)} = \alpha_e^{B(BB)} = \sqrt{3}/4 = 0.433013, \alpha_e^{A(AB)} = 0.406116,\alpha_e^{B(AB)} = 0.45453, \alpha_e^{A(BB)} = 0.378775$ and $\alpha_e^{B(AA)} =  0.473803$.

We can quantify the amount of order present in the packing by measuring the area under these ordered peaks. We study this in detail in Section \ref{MRJ_section}.
The disordered parts of the edge triangle areas, namely the areas not under any ordered peaks, exhibit characteristics similar to those of the contact angles (see Fig. \ref{prob_angle_figure}). This can be understood from the fact that the areas of the edge triangles are built from the underlying contact vectors and contact angles and that at low energies the contact vectors take well defined values.

\begin{figure}
\hspace{0.5cm}
\includegraphics[height=.35\columnwidth,angle=0]{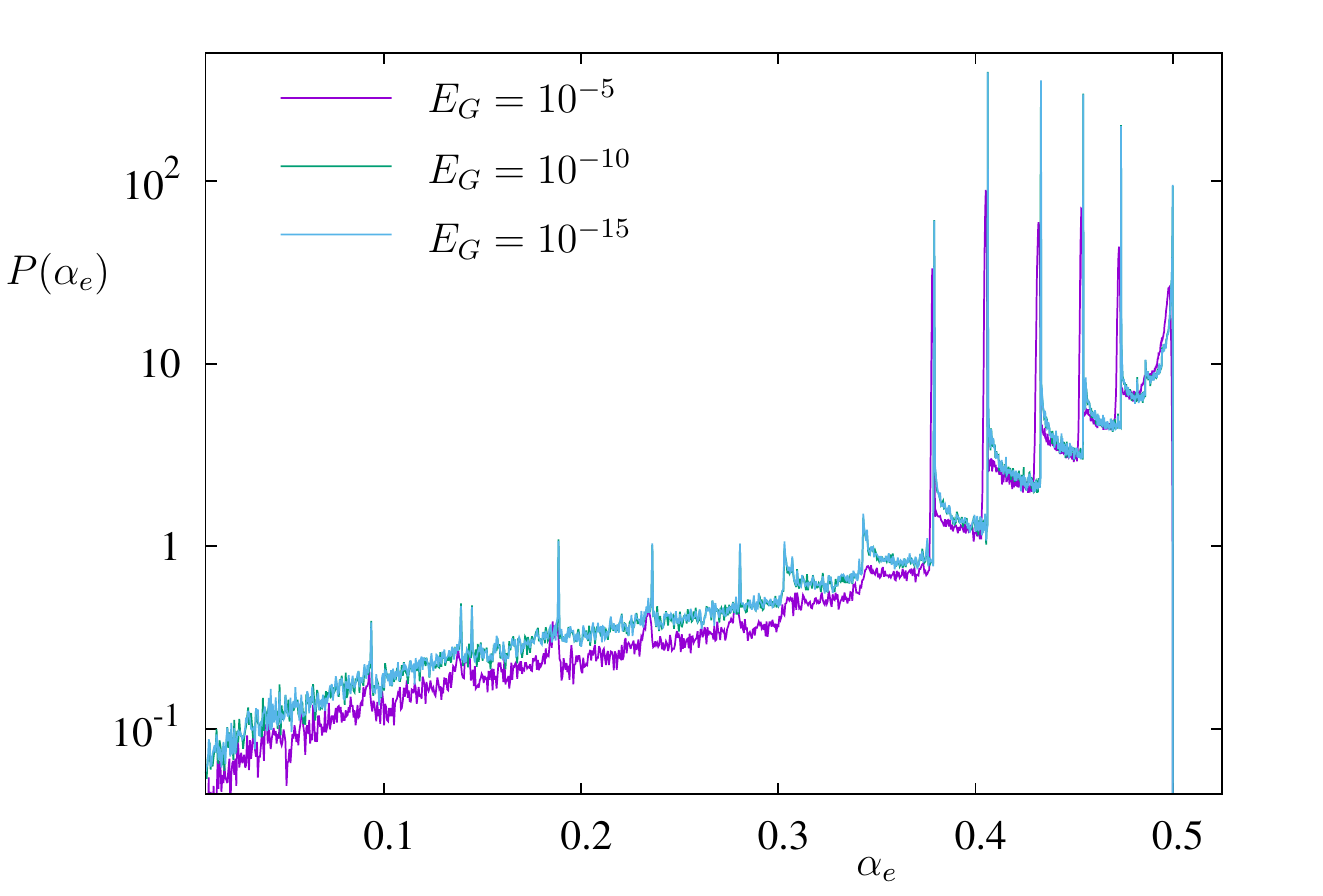}
\includegraphics[height=.35\columnwidth,angle=0]{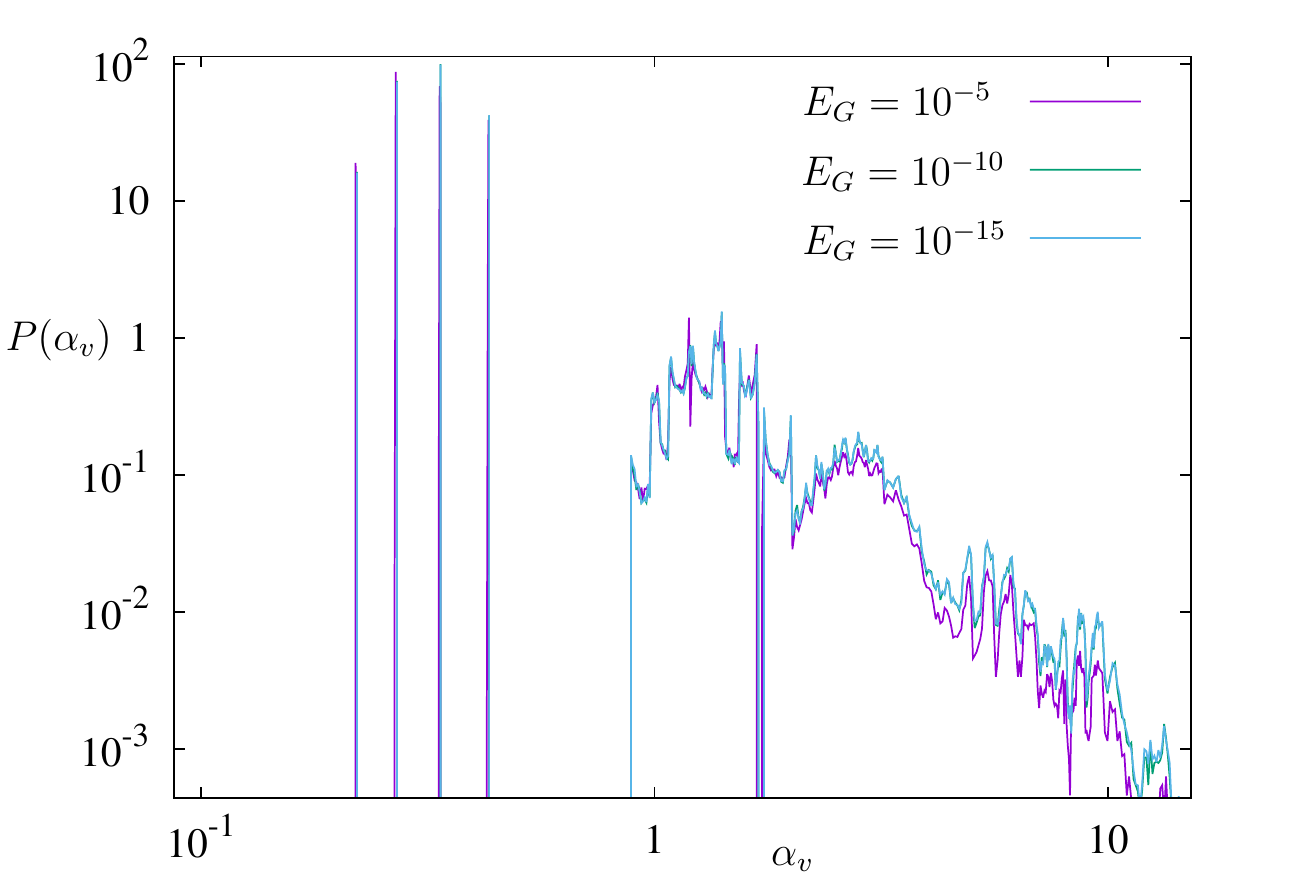}
\caption{{\bf (Left)} Distribution of areas of the edge triangles. The plot shows the distribution of $\alpha_e = a_{g,e}/\sigma_g^2$ for $N_G = 2048$ at different energies. $\alpha_e \to 1/2$ corresponds to disks with relative contact angles close to $\pi/2$. We find well defined ordered peaks that can be used to estimate the amount of order in the system. The five largest peaks arise from ordered structures formed by three disks in contact (see Fig. \ref{three_disks_figure}). These ordered peaks get sharper as $E_G \to 0^+$.
{\bf (Right)} Distribution of areas of void polygons $\alpha_{v} = a_{v}/{\sigma_A}^2$, where $\sigma_A$ is the radius of grains of type-$A$ in the given configuration. The plot shows $P(\alpha_v)$ for configurations of bidispersed grains with $N_G = 2048$ at different energies. We find four well-defined ordered peaks that correspond to three-disk ordered structures (see Fig. \ref{three_disks_figure}). The area under these ordered peaks can be used to estimate the fraction of order present in the system $\Psi_{O} \approx 0.369(1)$ as $E_G \to 0^+$.}
\label{grain_void_figure}
\end{figure} 

\subsection{Void Areas}
\label{void_areas_section}
Another quantity that can be used to measure disorder at the microscopic scale are the voids that are formed in jammed packings.
In order to characterise the spatial structure of these voids, we measure the distribution of the areas of individual void polygons ($a_v$). In this respect, our polygonal construction has an advantage over previous measures of spatial anisotropy, since we can identify well defined sections of space that belong to voids alone.
In order to account for the different sizes of the grains and changes in grain diameters for different configurations, we measure the following normalized void areas $\alpha_{v} = a_{v}/{\sigma_A}^2$, where $\sigma_A$ is the radius of grains of type-$A$ (the larger of the two) in a given configuration.  We find that even though $\{ \sigma_g \}$ varies for different configurations and also for different energies, the distribution of $\alpha_v$ is reproducible and also displays self averaging properties. 
In Fig. \ref{grain_void_figure} we plot the distribution of these normalized void polygon areas $P(\alpha_v)$ for a system of $N_G = 2048$ grains at three different energies $E_G = 10^{-5}, 10^{-10}$ and $10^{-15}$ approaching the marginally jammed state.
As with the distribution of edge triangles, we find well defined ordered peaks that persist even as we approach the marginally jammed state. The first four peaks in this distribution can be identified as arising from triangular voids ($z_v = 3$) that arise from three-disk ordered structures. These four peaks arise from the four possible combination of type-$A$ and type-$B$ disks in the bidispersed system, namely $(AAA)$, $(AAB)$, $(ABB)$, and $(BBB)$. At the  marginally jammed state with zero overlaps, the distances between the disks in contact can take only unique values  (see Fig. \ref{three_disks_figure}) leading to a unique values of the areas of these three sided void polygons. We can compute the positions of these peaks exactly. For the ratio of diameters $1:1.4$ these are $\alpha_v^{(AAA)} = 0.433013, \alpha_v^{(AAB)} = 0.33843,\alpha_v^{(ABB)} = 0.270553$, and $\alpha_v^{(BBB)} = 0.220925$.
In states with larger energies, the finite overlaps between disks causes a broadening of these peaks. Although the value at which the peaks occur are easily understood, the question of how many of the voids in the system contribute to them is non-trivial. 
In this context,
an interesting quantity to measure is the area under these four ordered peaks, we measure this in Section \ref{MRJ_section}.
The disordered parts of the void polygon areas exhibits interesting modulations, which can be attributed to ordered structures formed by four or more disks in contact.

\subsection{Maximally Random Jammed States}
\label{MRJ_section}
The marginally jammed states of soft disks have exactly zero overlap and can therefore be mapped on to systems with infinitely hard interactions. Such packings of frictionless hard disks have been the subject of continued interest and have a rich history \cite{aste_book_2008}. Hard disk packings generated from random protocols have recently been argued to be {\it maximally random} \cite{torquato_pnas_2014}, i.e. they present the least amount of order amongst all possible states available to the system.
A natural question to ask is then: how random are the states of frictionless soft disk packings as the unjamming transition is approached. Inversely we can study the amount of order present in these soft disk packings close to the marginally jammed state in order to understand the nature of such maximally random states.

A natural measure of disorder is the {\it absence} of the ordered structures formed by three disks in contact identified in Sections \ref{edge_triangles_section} and \ref{void_areas_section}.
In Section \ref{connectivity_section} we derived local conservation laws for the number of fictitious contacts, assigning them uniquely to the voids with $z_v \ne 3$.
We can then define a measure of disorder as simply the total number of fictitious contacts in the system, or alternatively as a density measure
\begin{equation}
\psi_{DO} = \frac{n_f}{N_V} ~~~\textmd{and}~~~ \Psi_{DO} = \sum_{v} \psi_{DO},
\label{disorder_measure_definition}
\end{equation}
where the disorder density $\psi_{DO}$ is assigned to every void.
The normalization ensures that $0 \le \Psi_{DO} \le 1$ as $N_G \to \infty$ (see Eqs. (\ref{void_bound}) and (\ref{fictitious_bound})).
Numerically, this can be easily computed for a given packing using relations derived in Section \ref{coordination_section}. As we have shown, the number of fictitious contacts attains its maximum value at isostaticity. 
Therefore in this measure isostatic packings naturally have the largest amount of disorder $\Psi_{DO} = 1$. 
Since we have also argued that marginally jammed packings that do not possess special conservation laws are isostatic, randomly created marginally jammed states are therefore the most disordered.
Similarly the hexagonally close packed structure has the least amount of disorder $\Psi_{DO} = 0$. 

\begin{figure}
\hspace{0.5cm}
\includegraphics[height=.35\columnwidth,angle=0]{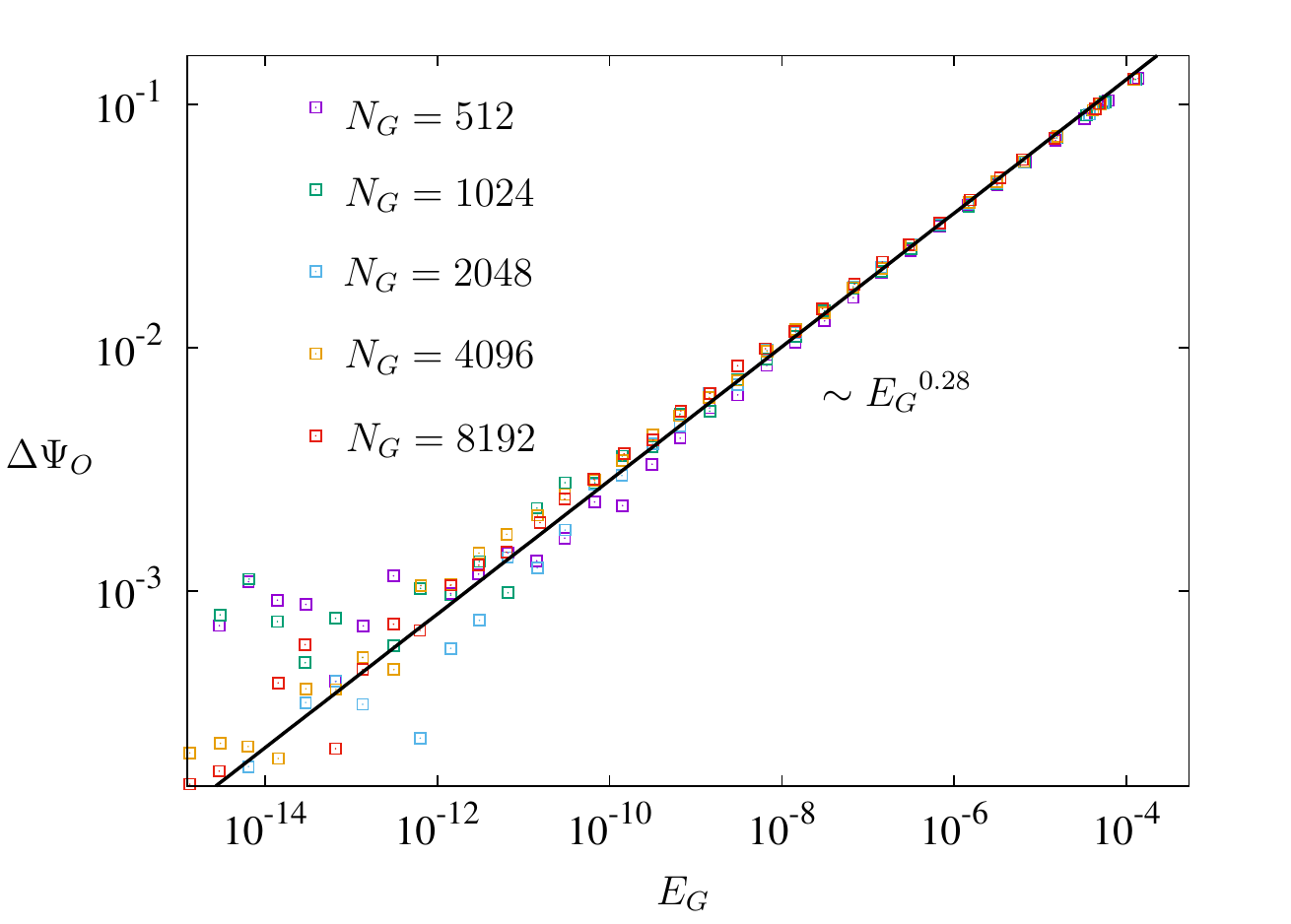}
\includegraphics[height=.35\columnwidth,angle=0]{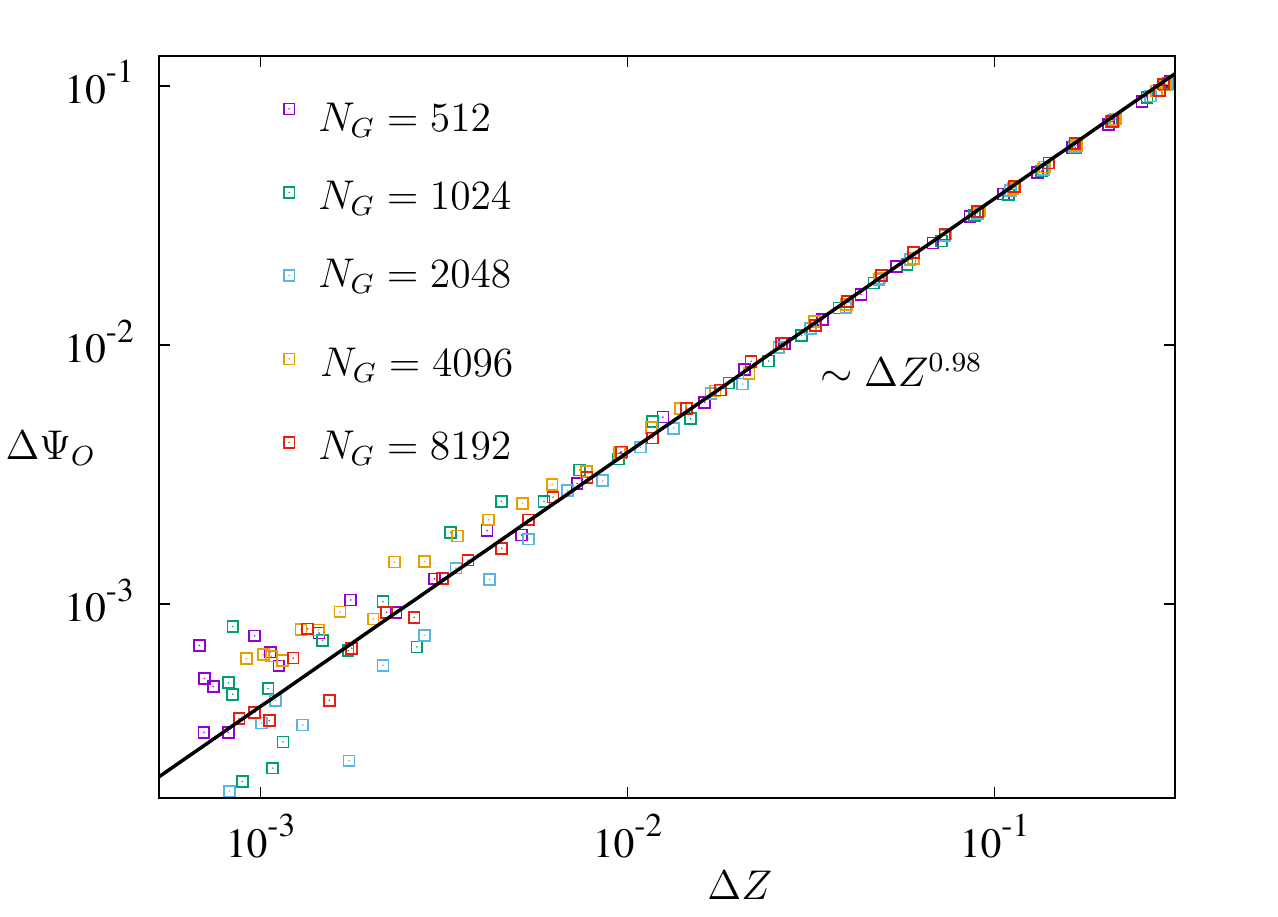}
\caption{{\bf (Left)} Scaling of the excess order in the system $\Delta \Psi_O$ with total energy per particle $E_G$. We find that the excess order scales as a power of the total energy in the system with exponent $\approx 0.28(3)$. {\bf (Right)} Scaling of the excess order $\Delta \Psi_{O}$ with the excess coordination $\Delta Z$. We find that the excess order scales as a power of $\Delta Z$ in the system with exponent $\approx 0.98(3)$.}
\label{psi_with_EG_figure}
\end{figure} 

Another closely related measure of {\it order} is the fraction of the system under the ordered peaks in the microscopic distribution functions measured in Sections \ref{edge_triangles_section} and \ref{void_areas_section}.
In the simplest construction we only account for the ordered structures formed by
three disks in contact in the system. This can be estimated from the amount of area under the five ordered peaks in the distribution of edge triangle areas or alternatively the four ordered peaks in the distribution of the void areas. This measure can be recognized as simply the total number of voids with a coordination $z_v = 3$. We can therefore define a measure of order as 
\begin{equation}
\psi_{O} = \frac{n_v(z_v = 3)}{N_V} = P(z_v = 3), ~~~\textmd{and}~~~ \Psi_{O} = \sum_{v} \psi_{O},
\end{equation}
where $n_v$ is the number of voids with connectivity $3$ in a packing with $N_V$ voids and $P(z_v)$ is the distribution of void connectivities studied in Section \ref{connectivity_section}.
As the peaks sharpen, $\Psi_{O}$ decreases, but remains finite even in the $E_G \to 0^+$ limit.
For example for configurations prepared with $N_G = 8192$ disks, at $E_G = 10^{-5}$ we find $\Psi_{O} \approx 0.435$, at $E_G = 10^{-10}$ we find $\Psi_{O} \approx 0.373$, and at $E_G = 10^{-15}$ we find $\Psi_{O} \approx 0.369$. Using finite size scaling as in Section \ref{grain_area_section}, we find that
\begin{equation}
\Psi_O \to \Psi_O^* = 0.369(1) ~~\textmd{as}~~ E_G \to 0^+  ~~\textmd{and}~~ N_G \to \infty.
\end{equation}
We obtain a similar estimate from the limit $\langle z_g \rangle \to \langle z_g \rangle_{\rm iso}$. Once again, as for the excess grain area, we can study the behaviour of the excess order $\Delta \Psi_{O} = \Psi_{O} - \Psi_{O}^*$ as the energy of the system and the coordination number is increased.
The scaling of the excess order with energy and $\Delta Z$ are shown in Fig. \ref{psi_with_EG_figure}. We find that $\Delta \Psi_{O}$ displays non-trivial scaling with energy and also with the excess coordination. 
\begin{eqnarray}
\nonumber
\Delta \Psi_{O} \sim (E_G)^{0.27(3)},\\
\Delta \Psi_{O} \sim (\Delta Z)^{0.98(3)}.
\end{eqnarray}
Intriguingly these are the same exponents as those for the scaling of the excess grain area. It would be very interesting to understand the origin of these scaling exponents in such disordered packings of soft disks.

\section{Correlations}
\label{correlation_section}

Although several probability distributions of local properties are reproducible and self averaging, this does not imply a completely uncorrelated behaviour on the microscopic level.
Several constraints such as force balance and the tilings constraints for polygons can cause local structures to emerge in correlated patterns. Hence it is important to assess the length scale over which these correlations persist in the real space networks, and specifically for the quantities that we have measured. In this section we compute network-based correlation functions that we use to assess the length scale up to which {\it short range order} persists in this system. We use these correlation functions to estimate the system sizes needed for self averaging behaviour to become valid.

\subsection{Particle and Contact Density Fluctuations}
In order to test the correlations in the system we measure the radial distribution function $P_{\rm radial}^{g}(r)$ defined as the probability that the centre of a disk $g'$ is within the annulus  $r$ and $r + dr$ centered around $g$, normalized by $2 \pi r$. Here we normalize the distances by the size of the grain from which it is measured $r = r_{g,g'}/\sigma_g$, where $r_{g,g'}$ is the distance computed in the real space packing $r_{g,g'} = |\vec{r}_{g'}-\vec{r}_{g}|$.
This is equivalent to computing a two point radial correlation function. We find peaks at well-defined distances which we can attribute to the short-range order present in the underlying packings. We have tested that the scaling near the ordered peaks follows the $1/\sqrt{r-1}$ behaviour observed in previous studies \cite{ohern_prl_2002, ohern_pre_2003}.

\begin{figure}
\hspace{0.5cm}
\includegraphics[height=.35\columnwidth,angle=0]{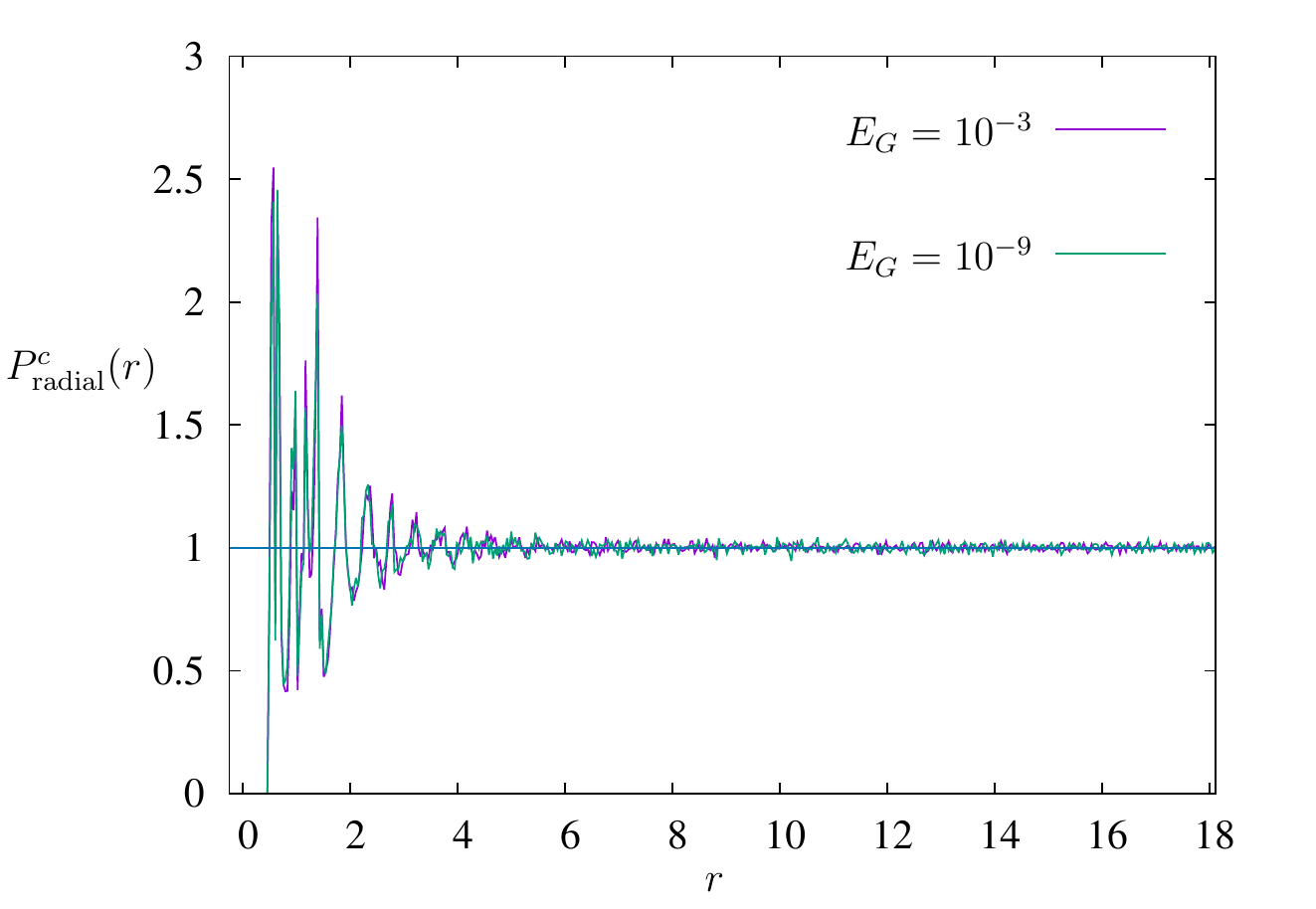}
\includegraphics[height=.35\columnwidth,angle=0]{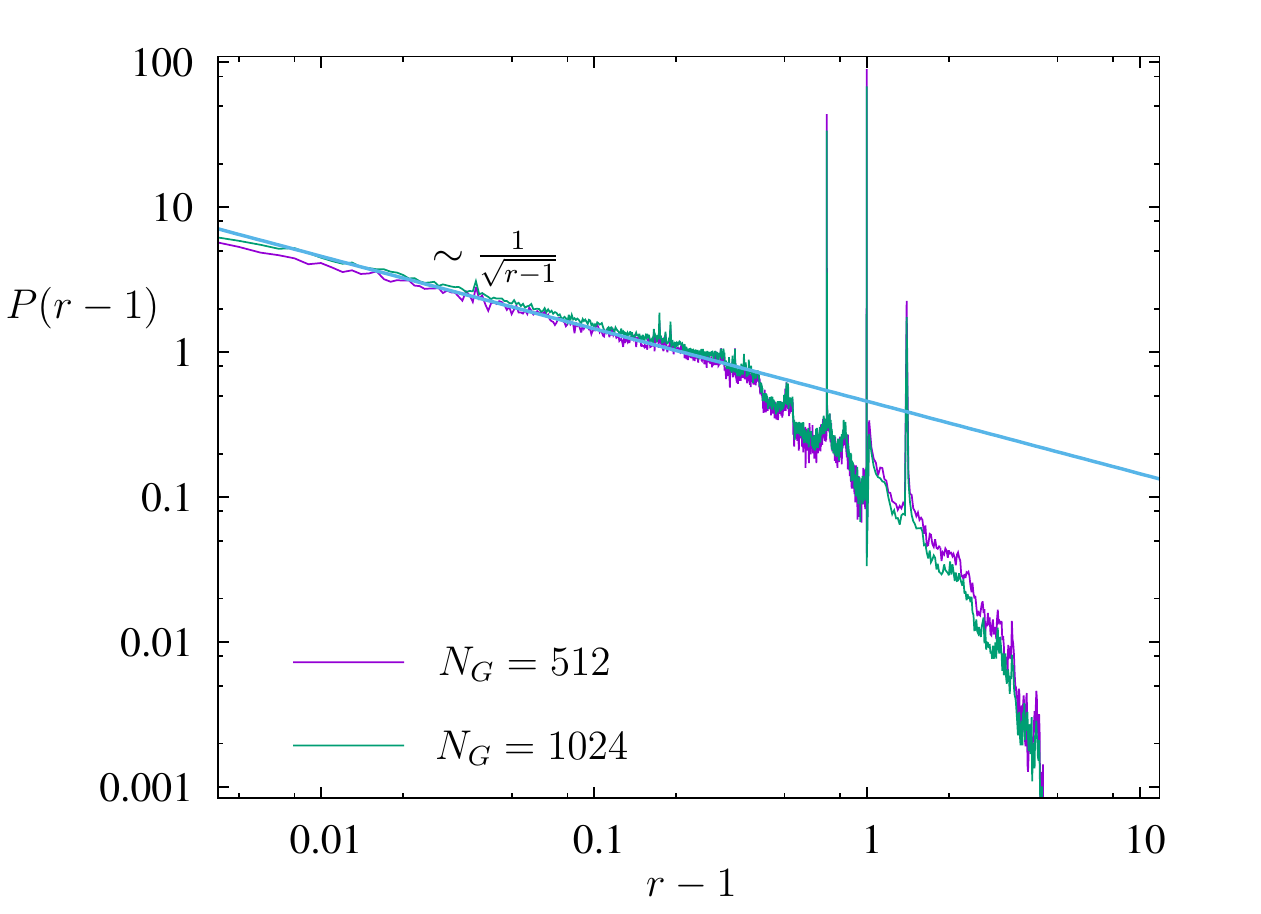}
\hspace{-1cm}
\caption{{\bf (Left)} The radial distribution function $P_{\rm radial}^{c}(r)$ for contacts, with $r = r_{c,c'}/\sigma_{g}$ and $r_{c,c'} = |\vec{r}_{c'}-\vec{r}_{c}|$. Here $g$ refers to the grain to which the central contact belongs. This has been computed for a packing with $N_G = 1024$ at different energies. We find well defined density modulations that are detectable up to $\sim 10$ grain diameters that has only a weak dependence on energy.
{\bf (Right)} Distribution of the length of fictitious contacts $P(r-1)$, where $r = |\vec{r}_{g,g'}|/(\sigma_g + \sigma_g')$ and $\vec{r}_{g,g'}$ is the length of the triangulation vector connecting grains $g$ and $g'$ with radii $\sigma_g$ and $\sigma_g'$ respectively. The plotted distribution is at the lowest measured energy $E_G = 10^{-15}$. We find a clear $\frac{1}{\sqrt{r-1}}$ behaviour as $r \to 1$. The drop off at $\sim 3$ grain diameters  sets the length scale for the linear dimension of the voids, independent of the system size.}
\label{correlations_figure}
\end{figure} 
 
Similarly we can define $P_{\rm radial}^{c}(r)$ as the probability that a contact $c'$ is within the annulus  $r$ and $r + dr$ centered around the contact $c$, normalized by $2 \pi r$. Once again we normalize the distances by the size of the grain to which the central contact belongs $r = r_{c,c'}/\sigma_g$, where $r_{c,c'}$ is the distance computed in the real space packing $r_{c,c'} = |\vec{r}_{c'}-\vec{r}_{c}|$. We use this distribution function to study the fluctuations in the number of {\it contacts} within a given radius of another contact. This is in effect a two point contact correlation function.
In Fig. \ref{correlations_figure} we plot $P_{\rm radial}^{c}(r)$ for a packing of $1024$ disks at varying energies. 
We notice that this radial contact distribution function exhibits periodic modulations that persist for $\sim 10$ grain diameters. This length scale is set by the underlying geometrical randomness and therefore has only a very weak dependence on the energy of the system.
This can be explained from the fact that contacts are located on the vertices of tilings of void and grain polygons. Since the distribution functions of the areas of these polygons exhibit ordered peaks, this also implies ordered peaks in the the linear size of these polygons. Next, since the distances between contacts can be formed by traversing linearly through these polygons, this leads to ordered peaks in the contact correlations. Since these distances are drawn randomly from the underlying distribution functions, these correlations will be randomized after a ``persistence length", $\xi \sim 10$ grain diameters in this case. This length scale sets a scale for the system sizes above which we can expect self averaging behaviour to become valid. Since we keep the box length fixed, the linear size (number particles traversed along a linear axis) of the system grows as $\sqrt{N_G}$. This suggests that the effect of these correlations can be neglected beyond systems with $N_G > \xi^2 \sim \mathcal{O}(10^2)$. This agrees with our finite size studies of the scaling behaviour near the unjamming transition.

\subsection{Length of Fictitious Contacts}
Finally, to understand the spatial correlations in the {\it triangulation network}, we measure the distribution of the lengths of fictitious contacts formed in the Delaunay triangulation graph. This gives us valuable information about the spatial separation of the disks that are almost in contact.
The Delaunay triangulation is a geometric spanner, i.e. the length $r$ of the shortest path along Delaunay edges is known to be  $\frac{\pi}{2} \le \textmd{max}\left[\frac{r}{r_e}\right] \le \frac{4\pi}{3\sqrt{3}}$, where $r_e$ is the Euclidean distance between vertices \cite{keil_gutwin_dcg_1992}. This allows us to understand the long distance correlations in the system based simply on the distributions of the lengths of vectors in the triangulation.
As we have shown, the number of fictitious contacts increases and attains its maximum value as the transition is approached.
This points to the fact that the fictitious contacts play a crucial role in determining the stability of the jammed networks close to the unjamming transition. Another important property of the fictitious contacts is that they traverse the void polygons, as shown in Section \ref{conservation_section}. Therefore the lengths of these vectors can be used to estimate the linear dimension of the voids. This is important in measuring quantities such as porosity, that has been of interest in the literature \cite{sastry_stillinger_pre_1997, guan_prl_2013, donev_stillinger_prl_2005}.

In order to understand the behaviour of the fictitious contacts near the transition, we measure the distribution function $P(r)$ defined as the probability that two grains are separated by a Delaunay edge of length $r$. We account for the different sizes of the disks by normalizing the length of the triangulation vectors as $r = |\vec{r}_{g,g'}|/(\sigma_g + \sigma_g')$ where $\vec{r}_{g,g'}$ is the length of the triangulation vector between grains $g$ and $g'$ and $\sigma_g$ and $\sigma_g'$ are the radii of the disks connected by the vector. Distances of $r > 1$ correspond to fictitious contacts and $r < 1$ correspond to real contacts. Since we have already studied the distribution of lengths of contact vectors in Section \ref{contact_vectors_section}, we focus on the $r > 1$ part of this distribution.
This differs from the standard pair correlation function \cite{silbert_liu_nagel_pre_2006} since this distribution function only measures nearest-neighbour distances on the Delaunay network.
In Fig. \ref{correlations_figure} we plot the distribution $P(r-1)$ for two system sizes with $E_G = 10^{-15}$, i.e. marginally jammed. 
We find a clear $\frac{1}{\sqrt{r-1}}$ behaviour as $r \to 1$, in agreement with previous studies of the pair correlation functions \cite{silbert_liu_nagel_pre_2006}. In this case as well, we find sharp peaks that we can identify as arising from the ordered structures formed by {\it four} disks in contact. Finally, we note that the sharp drop-off in the distribution at $\sim 3$ grain diameters sets the length scale for the linear size of the voids, independent of the system size.

\section{Discussion}
\label{conclusion_section}

In conclusion, we have introduced a network based framework for analysing  spatial characteristics of jammed packings. The polygonal construction differs from the well-known Voronoi measure of local volumes since it allows us to probe the internal structure of both grains {\it and} voids in jammed packings. We have found that the grain area serves as a reliable parameter with which to describe the unjamming transition. We found evidence for a well defined transition at an area fraction $A_G^* = 0.446(1)$. 
The construction of polygonal tilings allowed us to precisely study the scaling behaviour near the unjamming transition.
We measured the scaling properties of this area and found that it displays non-trivial scaling with both the energy and excess coordination as the transition is approached. We found new structural critical exponents $\beta_E = 0.28(2)$ and $\beta_Z = 1.00(1)$ that describe this scaling behaviour near the transition.
We expect the critical exponent $\beta_E$ to display non-universal behaviour for different potentials while the microscopic assignment of areas to the contacts makes the exponent $\beta_Z$ universal.
Our measured distribution functions revealed signatures of a finite order even in the marginally jammed state. We estimated this fraction of order using the amount of ordered structures formed by three disks in contact in the marginal state as $\Psi_{O}^* = 0.369(1)$. We also found that the excess order in the system displays non-trivial scaling near the transition, attaining a minimum value in the marginal state. Another interesting aspect of our analysis is the observed deviation of these exponents from simple fractions, similar to those observed in recent experiments \cite{brujic_arxiv_2016}. It would be very interesting to understand the origin of these non-trivial exponents in such systems.

A natural extension to this study would be to analyse the quantities studied in this paper at different values of polydispersity. Since packing fractions seem to display some degree of robustness to polydispersity near the transition, it would be interesting to test the sensitivity of the polygonal measure in this regard.
It would also be interesting to extend our analysis to convex particles with non-circular shapes, which our construction can straightforwardly be extended to. 
Our network construction is able to clearly distinguish between frictionless and frictional systems. An interesting aspect for further investigation would then be to study the effect of friction on the statistics of these jammed networks. Many of our constructions can be generalized to higher dimensional systems which would be an intriguing avenue of further research.
Finally, it would be very interesting to use the underlying distribution functions studied in this paper to construct microscopic theories that describe the scaling properties of such systems near the unjamming transition.

\section{Acknowledgements}
\label{acknowledgements_section}

We thank C. S. O'Hern, T. Bertrand and Q. Wu for providing configurations that were used in the initial analysis of this work. We acknowledge helpful discussions with D. Dhar. K. R. acknowledges helpful discussions with A. Narayanan.
This work has been supported by NSF-DMR 1409093 and the W. M. Keck Foundation.


\end{document}